\documentclass[sigconf]{acmart}

\makeatletter
\def\@ACM@checkaffil{
    \if@ACM@instpresent\else
    \ClassWarningNoLine{\@classname}{No institution present for an affiliation}%
    \fi
    \if@ACM@citypresent\else
    \ClassWarningNoLine{\@classname}{No city present for an affiliation}%
    \fi
    \if@ACM@countrypresent\else
        \ClassWarningNoLine{\@classname}{No country present for an affiliation}%
    \fi
}
\makeatother

\usepackage{amsmath,amsfonts}
 
\usepackage{amssymb}

\usepackage{stackengine}
\usepackage{scalerel}
\usepackage{xcolor}
\newcommand\dangersign[1][2ex]{%
  \renewcommand\stacktype{L}%
  \scaleto{\stackon[1.3pt]{\color{red}$\triangle$}{\tiny !}}{#1}%
}
\definecolor{darkgreen}{RGB}{0, 128, 0}

\usepackage{pifont} 
\usepackage{bbding} 

\usepackage{makecell}
\usepackage{colortbl} 

\usepackage{longtable}

\usepackage{lipsum}
\usepackage{tikz}
\usetikzlibrary{arrows.meta}
\usetikzlibrary{automata,positioning}

\usepackage{threeparttable}

\usepackage{fbox} 
\usepackage{fancybox}
\usepackage{framed}

\renewcommand*{\arraystretch}{1.5}%
\definecolor{tabred}{RGB}{230,36,0}%
\definecolor{tabgreen}{RGB}{0,116,21}%
\definecolor{taborange}{RGB}{250,124,30}%
\definecolor{tabbrown}{RGB}{171,70,0}%
\definecolor{tabyellow}{RGB}{251,253,169}%
\newcommand*{\vcorr}{%
  \vadjust{\vspace{-\dp\csname @arstrutbox\endcsname}}%
  \global\let\vcorr\relax
}%

\usepackage{mathtools}

\usepackage{lscape}
\usepackage[figuresright]{rotating}
\usepackage[graphicx]{realboxes}
\usepackage{adjustbox}
\usepackage{caption}

\usepackage{subfigure}

\usepackage{colortbl} 
\usepackage{xfrac}

\usepackage{appendix}
\usepackage{float}

\usepackage{multirow}

\def\BibTeX{{\rm B\kern-.05em{\sc i\kern-.025em b}\kern-.08em
    T\kern-.1667em\lower.7ex\hbox{E}\kern-.125emX}}
    
\usepackage{CJKutf8}  
\usepackage{pgfplots}
\usepackage{filecontents}

\usepackage{hyperref}

\usepackage{tabularx,makecell, caption}
\usepackage{enumitem} 

\newcolumntype{L}{>{\arraybackslash}X}

\usepackage{multicol}
\usepackage{listings}
\usepackage{blindtext}
\usepackage{etoolbox,xstring,mfirstuc,textcase}
\usepackage{listings}

\usepackage{subfiles}
\usepackage[most]{tcolorbox}

\usepackage{pifont}
\usepackage{ulem} 
\newcommand{\hlhref}[2]{\href{#1}{\textcolor{blue}{\underline{#2}}}}

\definecolor{armygreen}{rgb}{0.29, 0.33, 0.13}

\usepackage{framed}
\definecolor{Gray}{gray}{0.9}
\definecolor{shadecolor}{gray}{0.95}
\usepackage[most]{tcolorbox}

\usepackage{xcolor}
\theoremstyle{plain}         
 
\newtheorem{thm}{Theorem}    
  
\newtheorem{defi}{Definition}
\theoremstyle{definition}

\newtheorem*{prf}{Proof}

\usepackage{siunitx}
\usepackage{comment}
\usepackage{indentfirst} 
\usepackage{framed} 

\usepackage[font=small,labelfont=bf,tableposition=top]{caption}
\usepackage{booktabs}
\usepackage{dashbox}

\usepackage{listings}
\usepackage{url}

\usepackage{color}

\usepackage[ruled,linesnumbered]{algorithm2e}
\usepackage{algpseudocode}
\usepackage{multicol}

\renewcommand\footnotetextcopyrightpermission[1]{} 

\begin{document}
\title{Maximizing NFT Incentives: References Make You Rich}


\author{Guangsheng Yu$^{1}$$^\star$, Qin Wang$^{1}$$^\star$, Caijun Sun$^{2}$, Lam Duc Nguyen$^{1}$, \\  H.M.N. Dilum Bandara$^{1}$, Shiping Chen$^{1}$}\thanks{$^\star$ These authors contributed equally to the work.}

\affiliation{
\textit{$^1$CSIRO Data61, Australia}\\
\textit{$^2$Zhejiang Lab, China} 
}


\begin{abstract}
In this paper, we study \textit{how to optimize existing Non-Fungible Token (NFT) incentives}. 
Upon exploring a large number of NFT-related standards and real-world projects, we come across an unexpected finding. That is, the current NFT incentive mechanisms, often organized in an isolated and one-time-use fashion, tend to overlook their potential for scalable organizational structures.

We propose, analyze, and implement a novel \textit{reference incentive} model, which is inherently structured as a Directed Acyclic Graph (DAG)-based NFT network. This model aims to maximize connections (or references) between NFTs, enabling each isolated NFT to expand its network and accumulate rewards derived from subsequent or subscribed ones. We conduct both theoretical and practical analyses of the model, demonstrating its optimal utility. 

\end{abstract}


\maketitle


\section{Introduction}\label{sec-intro}

Non-Fungibie Tokens (NFTs) made a big splash in mid-2021~\cite{wang2021non} and have since become a major player in the market. The peak week saw a whopping \$3.24b in NFT transactions (22/8/2022-29/8/2022, $\mathsf{TheBlock}$\footnote{Data sources by $\mathsf{TheBlock}$ \url{https://www.theblock.co/data/nft-non-fungible-tokens/nft-overview}; $\mathsf{Dune}$ \url{https://dune.com/hanxilgf/nft-overview}; and \url{https://dune.com/hildobby/NFTs}; $\mathsf{DappRadar}$ \url{https://dappradar.com/rankings/nft/marketplaces}.\label{ft:data}}). Recently, the 
average daily trading volume has reached 
\$36m, attracting over 14k active daily traders (Jan 2024, $\mathsf{Dune}$\footref{ft:data}). Originating in the Ethereum ecosystem, NFTs have spread across mainstream platforms like Solana, BNB chain, Polygon, Avalanche, Optimism, and Rarible ($\mathsf{DappRadar}$\footref{ft:data}). Trading is bustling on specialized platforms such as Blur, OpenSea, OKX-NFT, Tensor, Reservoir, LooksRare, X2Y2, and Sorare. Even the previously underestimated Bitcoin ecosystem~\cite{binance1}, leveraging the inscription technique~\cite{li2024bitcoin}, has stepped into the NFT arena and introduced new NFT products like BRC20 and Ordinals, finding success in mid-2023~\cite{wang2023understanding}.

ERC-721~\cite{erc721} has emerged as the de facto standard for NFTs,
accounting for the majority (98.8\%, $\mathsf{Dune}$), and the remaining 1.2\% 
mostly follow the ERC-1155~\cite{erc1155} standard. These standards play a pivotal role in establishing the
core functionalities of NFTs, namely, \textit{minting} and \textit{transferring}. The \textit{minting} process involves the creation of a new NFT, and it is typically executed through a smart contract on the blockchain. During minting, unique metadata (e.g., name, description, and associated media file) are defined. Once minted, NFTs are owned (cryptographically protected by private keys) by individuals/entities and stored in digital wallets. The \textit{transfer} operation allows for the seamless 
change of ownership of NFTs between different wallets. Each transaction will be recorded on-chain, providing a transparent and immutable history of ownership.

\smallskip
\noindent\textbf{Earn money via NFTs?} Due to their digital uniqueness, NFTs have gained prominence in various industries, such as digital arts and games, enabling creators to monetize their digital assets. 
We consider the following ways of benefiting from NFTs. (i) A creator can create and sell 
NFTs by minting/listing them on marketplaces (e.g., CryptoPunks and Ape Yacht Club are the top collection list as per $\mathsf{DappRadar}$), which provides an efficient method for creators to generate income. 
(ii) NFTs can be traded in secondary markets, ideally by buying low and selling high based on market trends and popularity.  (iii) Owners can participate in play-to-earn (P2E)~\cite{yu2022sok} NFT games to earn/sell in-game NFT items (e.g., Stepn and Decentraland). (iv) NFTs can be staked in niche platforms (e.g., NFTX, BAND Royalty, and Polychain Monsters) for rewards. (v) NFTs can be leased (as specified in EIP-\hlhref{https://eips.ethereum.org/EIPS/eip-4907}{4907}\&\hlhref{https://eips.ethereum.org/EIPS/eip-5006}{5006}) to others via secondary markets or DApps (e.g., Renfter, Vera, IQ Protocol, and Unitbox) for use in games. Other indirect methods, such as investing in NFT-related companies, are excluded from this work.

\smallskip
\noindent\textbf{\textcolor{red}{\dangersign} In quest of a sustainable incentive.} However, these approaches typically concentrate on rewarding specific behaviors (e.g.,  minting, selling, trading, and staking) 
and are limited to generating static (\textit{one-time}) and fixed (\textit{predictable}) income, lacking the potential to establish a continuous revenue stream. Consequently, NFT users might be incentivized to focus on selling as many newly created NFTs as possible, aiming to extract the highest bid as long as the selling price exceeds  
transaction fees. This could result in the creation of numerous meaningless NFTs and transactions, leading to network congestion and high transaction fees. Moreover, such incentives contradict practical scenarios wherein Intellectual Property (IP) creators seek sustained benefits from their successful products. For instance, a singer consistently earns royalties each time her songs are used for commercial purposes. This discrepancy motivates our research question:

\smallskip
\textit{Is it feasible to devise an optimally sustainable incentive mechanism for NFTs, or broadly, for larger decentralized IP-related production?}
\smallskip

We present our efforts (i.e., \textbf{\textit{contributions}}) step by step.

\smallskip
\noindent\textbf{\ding{172} Newly identified challenges in incentivizing NFTs.} 
Building on the preceding narrative, we noticed a previously overlooked constraint that significantly impedes the growth of NFT incentives: their \textit{topology}. Presently, while NFTs may belong to a collection, each NFT is designed in isolation. Most existing NFT implementations (cf. Table~\ref{tab-tokenstandard}) develop their functionalities on an individual NFT basis (e.g., defining roles as per EIP-\hlhref{https://eips.ethereum.org/EIPS/eip-7432}{7432}, extending functions as in EIP-\hlhref{https://eips.ethereum.org/EIPS/eip-5308}{5308}, and adding parameters as per EIP-\hlhref{https://eips.ethereum.org/EIPS/eip-5007}{5007}) without direct connections (or capabilities of being attached) to broader NFT-formed networks. The isolated nature of NFTs limits their effective relationships, resulting in their inability to establish a dynamic income flow over time. As a response, we dive into the intricacies of NFT networks to explore how connections are established.

\begin{figure}
    \centering
    \includegraphics[width=0.99\linewidth]{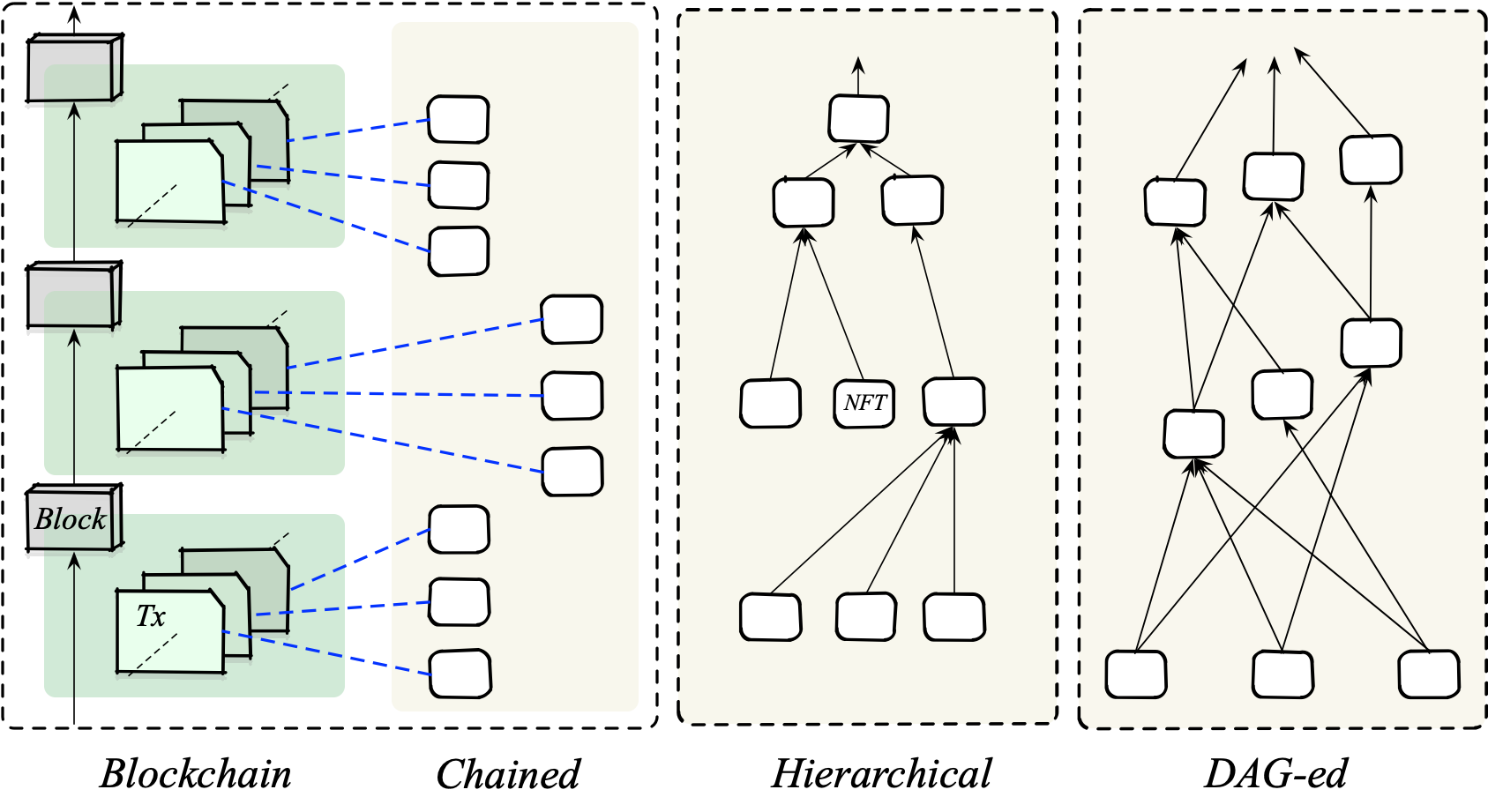}
    \caption{NFT formed network.}
    \label{fig:nfttopo}
    \vspace{-0.9em}
\end{figure}

\smallskip
\noindent\textbf{\ding{173} Discovering existing ways of structuring NFTs.} We have investigated mainstream NFT networks (see Fig.~\ref{fig:nfttopo}) based on all existing NFT-related standards (more in Table~\ref{tab-tokenstandard}). 

\begin{itemize}
    \item \textit{Chained design.} Each NFT is linked to a regular transaction. Despite the isolation of NFTs, an NFT retriever can find their history by searching the associated transactions. Notably, these connections are inherently formed by transactions in layer-one rather than direct links in L2\footnote{In the context of blockchain, layer-one (L1) represents the base or foundational layer of a blockchain network, including the main blockchain protocol and consensus mechanism. Layer-two (L2)~\cite{gudgeon2020sok} refers to solutions built on top of L1 blockchains, consisting of both scaling technologies (e.g., sidechains, state channels, and off-chain protocols) and upper-layer decentralized applications (DApps).}.
    
    \item \textit{Hierarchical design}. The network adopts a structured hierarchy (as exemplified in EIP-\hlhref{https://eips.ethereum.org/EIPS/eip-6150}{6150}), where each level signifies a distinct degree of abstraction or specialization, akin to scenarios in file storage. Ultimately, the network's topology in L2 takes on a tree-based shape.
    
    \item \textit{DAG design.} The network forms a \textit{directed acyclic graph} in L2, where an NFT may reference multiple ancestor NFTs or be referenced by multiple subsequent ones (EIP-\hlhref{https://eips.ethereum.org/EIPS/eip-5521}{5521}).
    
\end{itemize}

\smallskip
\noindent\textbf{\ding{174} Establishing a uniform \textit{reference incentive} model}. 
It is evident that a DAG-based topology maximizes NFT connections, allowing for increased references. The more references an NFT builds, the greater its potential to generate or extract revenues becomes. 
To step further, we accordingly formulate a generic NFT incentive model, which we refer to as the \textit{reference incentive model}. Our model follows the straightforward recognition that is calculated by a participant's net income, i.e., \textit{payoff function} as $\mathcal{U}$ = \textit{income} as $\mathcal{I}$ -  \textit{outcome} as $\mathcal{O}$ (see Sec.\ref{subsection-incentivemodel}). However, due to the inherent complexity of the network, we confront a series of difficulties.

\smallskip
\noindent\textit{\underline{Technical challenge (TC)-I:} How to accurately measure complex connections or references over time?}

\smallskip
\noindent\textbf{\ding{175} Reply TC-I:
Crafting and refining the model} (Sec.\ref{subsec-outcome}-\ref{subsec-payoff}). 
In formulating the income aspect, our principles center around evaluating two key dimensions: \textit{reference connections} and \textit{stateful accumulator}. Regarding the former, we introduce two novel metrics tailored to the formed DAG network: participation scale (\textit{horizontal} level, count by \textit{in-degree} and also referred to as the reference list $\vec{\Theta}$) and depth (\textit{vertical} level, represented as $d$). These metrics serve to gauge the potential engagement within the NFT network. For the former, an NFT publisher might have numerous followers (consider, for instance, a hit song remixed by multiple direct producers). As for the latter, we examine the publisher's continuous income from an accumulated standpoint. This perspective accounts for an NFT created by the publisher having repetitive references over time (e.g., a hit song remixed \textit{successively} by multiple producers).

Regarding the outcome function (equiv. cost function), we incorporate a range of parameters, such as descending rate $\sigma$ and interest rate $q$ (see Sec.\ref{subsec-outcome} for details), to establish constraints that prevent perpetual payments for any NFT, regardless of its value. Our aim is for the value of a published NFT to appropriately diminish with increasing depth. 
This approach ensures that a newcomer to the market can engage with a \textit{stepwise and fair} pricing structure: paying a premium for a recently popular NFT and less for an older one, irrespective of its classic status\footnote{We operate under the assumption that the majority of IP products will experience a natural decline in value over time. Exceptional cases such as rare artworks, which tend to appreciate in value, are not considered within the scope of this analysis.}. Importantly, the profit distribution and charging process can be concluded within a finite number of rounds, akin to the concept that a patent becomes free after a certain period.

Our proposed incentive model offers several additional advantages: (i) being tailored to diverse strategies, with every parameter being adaptable; (ii) facilitating the dynamic joining and departure of subscribed users over time, as well as providing a real-time cumulative revenue reflection; and (iii) being versatile for various incentive cases (either layer-two products or layer-one consensus incentives) as long as their network structures involve multi-agents and are formed in DAG. 

However, the intricate reference dynamics between publishers and followers pose difficulties in analyzing their best incentives.

\smallskip
\noindent\textit{\underline{Technical challenge (TC)-II:} Does the optimal utility exist?}
\smallskip

\noindent\textbf{\ding{176} Reply TC-II: Theoretical analyses via game theory} (Sec.\ref{sec-ananlysis}). 
We provide a primary theoretical analysis via game theory (in particular, a Stackelberg game~\cite{fang2021introduction}) to determine the optimal utility for users in decentralized services that is anchored on three key theorems: \textit{finality}, \textit{complexity}, and \textit{solvability}. 

Concretely, \textit{finality} ensures distribution processes within a finite number of rounds, crucial for predictable and reliable transactions. \textit{Complexity} shows the non-convex and NP-hard nature of optimizing the payoff function, highlighting the intricate challenges in distributed environments. \textit{Solvability} proves that users can eventually achieve optimal outcomes (namely, mixed strategy Nash Equilibrium) in competitive, non-cooperative settings.

These theorems collectively demonstrate the existence of an optimal utility for game players. However, obtaining such a corresponding solution is highly complex (computationally infeasible), indicating its impracticality for everyday users. We progress in different ways by leveraging AI-empowered solutions to maximally approach the optimal solution.

\smallskip
\noindent\textit{\underline{Technical challenge (TC)-III:} What could be the practical method for approaching the optimal utility?}
\smallskip

\noindent\textbf{\ding{177} Reply TC-III: Practical training via deep reinforcement learning (DRL) approaches} (Sec.\ref{sec-drl}-\ref{sec-experiment}).
Building on TC-II's response, we highlight the advantages of using DRL for complex optimization challenges. DRL excels in high-dimensional spaces and interactive learning, offering adaptability in dynamic scenarios. Its strength lies in balancing strategy exploration and exploitation, learning intricate policies that traditional methods may not capture. We correspondingly define key DRL elements (agents, environment, episodes, states, actions, policies, and reward functions) to simulate the actual NFT competitive scenarios  (Sec.\ref{subsec-drl-model}) and present its detailed approach to training and operating (Sec.\ref{subsec-drl-solve}). 

However, given that our reference incentive model allows for an unrestricted number of references (\textit{out-degree}) in each round, we need to streamline the training task for practical implementation.

We accordingly select a machine-learning (ML) resource marketplace~\cite{9796833,9445602} as our focus (Sec.\ref{sec-experiment}), which (i) limits references to two (\textit{out-degree} set to 2): one representing a \textit{dataset} and the other a \textit{model}\footnote{Represented as URLs when embedding to NFT fields.}; (ii) aptly mirrors the interest in ML markets during the era of AI, offering a viable case study for decentralized intelligence.

Our practical use case presents a rich observation, wherein:

\begin{itemize}
    \item The system skillfully navigates NFT quality diversity, ensuring fairness amidst varying distribution patterns.

    \item Economic elements, notably interest rates, influence publisher tactics, highlighting the need for balanced settings in the evolving NFT market.

    \item Increasing publisher diversity and adjusting NFT settings foster fairer reward distribution, essential for a competitive NFT environment.
\end{itemize}

Further, we present further discussions on extended capabilities (Sec.~\ref{subsec-capability}) and potential applications (Sec.\ref{subsec-application}) of our incentive model to existing scenarios in different dimensions.

\smallskip
\noindent\textcolor{teal}{\ding{171}} A very short summary of our deliveries in this study:

\begin{itemize}
    \item New \textit{structuring} bottleneck in NFT incentives;
    \item A generic \textit{referece incentive} model;
    \item Proved to be game-theoretically optimal;
    \item Proved to be practically solvable via DRL;
    \item Evidence from an implemented ML-based use case.
\end{itemize}

\section{The Reference Incentive Model}\label{sec-model}

Our study concentrates on the incentive analysis within the established NFT (DAG) network (cf. \textit{left} in Fig.\ref{fig:nfttopo}). In this section, we present the reference incentive model with theoretical analyses.

\subsection{Our Reference Incentive Model}\label{subsection-incentivemodel}

\noindent\textbf{Design principles}. In our model, an NFT publisher may act as both a \textit{leader} and a \textit{follower}: they are the leader of the resources they provide and a follower of other publishers' resources. Each NFT, positioned as a node on the DAG\footnote{Such an NFT network can be formed via EIP-\hlhref{https://eips.ethereum.org/EIPS/eip-5521}{5521}.}, signifies a certain type of unique resource (e.g., instantiated as \textit{dataset} or \textit{model} in Sec.\ref{sec-experiment}). The directed edges of the graph embody the reference relationship among NFTs. Intuitively, the incentive mechanism should operate under three orthogonal guiding principles:

\begin{itemize}
\item[\textbf{P1:}] An NFT with a wider referencing bandwidth (horizontally, count by \textit{in/out degree}) should receive higher incentives.

\item[\textbf{P2:}] An NFT with a deeper referencing depth (vertically, count by \textit{depth}) should accumulate more revenue.

\item[\textbf{P3:}] The mechanism employs a weighted system for profit distribution to ensure distribution is completed within a finite number of rounds.
\end{itemize}

\smallskip
\noindent\textbf{Utility function.} For any participant, his/her utility (equiv. \textit{payoff function}) is straightforwardly calculated by the gap between revenues (\textit{income}) and costs (\textit{outcome}). From this, we formally define the payoff function $\mathbb{U}$ for each NFT, with $\mathcal{I}$ representing the accumulated income and $\mathcal{O}$ symbolizing the cost associated with minting this NFT. To simplify the explanation, we omit the subscripts from the notations while concentrating on one particular NFT $\mathcal{R}_{i,h,\Theta}$. More parameters are defined in Table~\ref{parameter_model}.

\begin{defi}
The payoff function $\mathbb{U}$ of a publisher $j$ minting $i$-th NFT $\mathcal{R}_{i,h,\Theta}$ at the block height $h$ with a reference list of $\Theta$ is given by
\begin{equation*}
\mathbb{U}^j({\mathcal{R}_{i,h,\Theta}})= \mathcal{I}^j_{\mathcal{R}_{i,h,\Theta}} - \mathcal{O}^j_{\mathcal{R}_{i,h,\Theta}}.
\end{equation*}
\end{defi}

We then present how to construct the cost function $\mathcal{O}^j_{\mathcal{R}_{i,h,\Theta}}$ and the income function $\mathcal{I}^j_{\mathcal{R}_{i,h,\Theta}}$. For simplicity, we omit the index $j$ unless differentiating between multiple publishers is necessary.

\subsection{Outcome Function} \label{subsec-outcome}
$\mathcal{O}_{\mathcal{R}_{i,h,\Theta}}$ denotes the costs borne by any NFT publisher, where a unique NFT, $\mathcal{R}_{i,h,\Theta}$ (minted at the $i$-th order during round $h$), has an initial minting fee of $p_0$. The calculation for $p_0$ is as follows:

\begin{equation*}
\begin{aligned}
p_0
= & \hat{\mathcal{O}}\times\lbrack w_0, w_1, w_2, \dots, w_{|\vec{\Theta}|}\rbrack, \text{ with}\sum_{i=0}^{|\vec{\Theta}|} w_i=1 \text{ and }w_i \in \mathbb{R}_0^+,
\end{aligned} 
\end{equation*}
\textit{here, $p_0$ includes a list of ratios, where $d$ previous NFTs are denoted, and $w_i$ represents the profit-sharing ratio/weight for the $i$-th NFT in the reference list $\vec{\Theta}$.} 
Note that allocating the values $w_0$ requires an initial fixed expense $\hat{\mathcal{O}}>0$ for releasing each NFT, independent of the size of the reference list $\vec{\Theta}$. To make it clear, we further denote $w_0$ as the weight of \textit{self-references} with $w_i$ being the weights of all other \textit{cross-references} $\forall i>0$. 

\smallskip
In practice, $p_0$ can be further expanded to:
\begin{equation*}
\begin{aligned}
p_0
= & \left(\hat{\mathcal{O}} + \mathcal{O}\right)\times\lbrack w_0, w_1, w_2, \dots, w_{|\vec{\Theta}|}\rbrack, 
\end{aligned} 
\end{equation*}
\textit{where $\mathcal{O}$ is a variable top-up fee, acting as an additional payment to those who previously contributed and own particular NFTs that the new NFT refers to.}

\smallskip
The publisher has the choice to immediately pay a portion of the total fee, defined as the down-payment ratio $\lambda$, while the remaining is spread across the next $\hat{d}$ rounds, with $\hat{d} = \phi(1-\lambda)$, where $\phi$ is a constant function. Additionally, an extra but optional payment $\lbrace\pi_{r}: 0 \leq\pi_{r}\leq \bar{\pi_r}\rbrace$ can be done by any publisher who is about to publish a new NFT. This payment $\pi_r$ is linked to the following parameters by adjusting them accordingly to reap more profits:

\begin{itemize}
    \item \textit{Decay parameter ($d$)}: The decay parameter $d$ now becomes $\hat{d} + \Delta d$, where $\hat{d}$ is the initial decay parameter, and $\Delta d = f(\pi_r),\Delta d\in\mathbb N$, with $f$ being a monotonically increasing function. Therefore, a larger $\pi_r$ leads to a larger decay parameter $d$, extending the valid rounds of profit distribution.

    \item  \textit{Interest rate ($q$)}: The interest rate $q$ is modified from the initial interest rate $\hat{q}$ based on the payment $\pi_r$. We define this relationship as $(1+q) = g(\hat{q}, \pi_r)$, where $g$ is a monotonically decreasing function in $\pi_r$. This means a higher $\pi_r$ leads to a lower interest rate.

    \item  \textit{Descending rate ($\sigma$)}: The descending rate $\sigma$ is adjusted from the initial descending rate $\hat{\sigma}$, according to the payment $\pi_r$. We define this as $\sigma = l(\hat{\sigma}, \pi_r)$, where $l$ is a monotonically decreasing function in $\pi_r$. Thus, a larger $\pi_r$ results in a lower descending rate.
\end{itemize}

\begin{table}[t]
{\color{black}
\renewcommand\arraystretch{1}
\caption{Notation Definition}\label{tal: notation}
\begin{center}
\vspace{-0.5em}
\begin{threeparttable}
\resizebox{\linewidth}{!}{
\begin{tabular}{c|p{6.9cm}}
\toprule
\textbf{Notation} & \textbf{Definition} \\
\midrule
$N$ & The number of NFT publishers in the network.\\
$\vec{\Theta}$  & The reference list generated by referrable links.\\
$\lambda$ & The down-payment ratio of minting a new NFT.\\
$\vec{w}$ & The list of weights assigned to each referring edge in a reference list $\vec{\Theta}$. \\
$d$ & The measure of the maximum depth in the reference chain for a particular NFT, a.k.a., the decay parameter. \\
$q$ & The interest rate in outcome with increasing depth towards the decay parameter $d$. \\
$\sigma$ & The descending rate in income with increasing depth towards the decay parameter $d$. \\
$p_0$ & The one-off payment to each owner of $\vec{\Theta}$ if a publisher decide not to pay by installment.\\
$\pi_r$ & The optional payment in outcome that can be in favor of a longer $d$, a lower $q$, and a lower $\sigma$.\\
$\varepsilon$ & The probability that an NFT $\mathcal{R}_{i,h,\Theta}$ gets referred by a new NFT $\mathcal{R}'_{i,h,\Theta} \neq \mathcal{R}_{i,h,\Theta}$.\\
$\mathcal{R}_{i,h,\Theta}$ & The $i$-th referrable NFT minted at round $h$ with $\vec{\Theta}$. \\
$\mathcal{O}_{\mathcal{R}_{i,h,\Theta}}$ & The accumulated cost of publishing an NFT $\mathcal{R}_{i,h,\Theta}$. \\
$\mathcal{I}_{\mathcal{R}_{i,h,\Theta}}$ & The accumulated income of publishing an NFT $\mathcal{R}_{i,h,\Theta}$. \\
$\Omega_{\mathcal{R}_{i,h,\Theta}}$ & The NFTs candidate set for the new NFT $\mathcal{R}_{i,h,\Theta}$ to be selected, forming the reference list $\vec{\Theta}$.\\
$\hat{\mathcal{I}}$ & The initial fixed reward granted to publishers if the performance outperforms the entire $\Omega_{\mathcal{R}_{i,h,\Theta}}$.\\
$\hat{\mathcal{O}}$ & The initial fixed expense to publish a valid NFT.\\
\bottomrule
\end{tabular}
}
\end{threeparttable}
\end{center}
\label{parameter_model}
}
\vspace{-0.2in}
\end{table}%



The cost function ($\mathcal{O}_{\mathcal{R}_{i,h,\Theta}}$) with compound interest, which now depends on the payment $\pi_{r}$, is as follows:

\begin{equation*}
\begin{aligned}
\mathcal{O}_{\mathcal{R}_{i,h,\Theta}} 
= & \lambda p_0 + \pi_r + (1+q)^{1-\varepsilon} \frac{p_0 (1-\lambda)}{d} + \\
& (1+q)^{2-\varepsilon} \frac{p_0 (1-\lambda)}{d}   + ...  +  (1+q)^{d-\varepsilon} \frac{p_0 (1-\lambda)}{d} \\ 
= & \lambda p_0 + \pi_r+ \sum_{j=1}^{d} g(\hat{q},\pi_r)^{j-\varepsilon}  \frac{p_0 (1-\lambda)}{d}, 
\end{aligned}
\end{equation*}

\begin{equation*}
\begin{aligned}
\mathcal{O}_{\mathcal{R}_{i,h,\Theta}}
= & \begin{cases}
\lambda p_0 + \pi_r+ \sum_{j=1}^{d} g(\hat{q},\pi_r)^{j-\varepsilon} \frac{p_0 (1-\lambda)}{d}, & \text{if}\ \lambda \neq 1 \\
p_0  & \text{if}\ \lambda = 1
\end{cases}
\end{aligned}
\end{equation*}
\textit{where the parameter $\varepsilon$ represents the probability that this NFT gets referred when every time there appears a new NFT being published onto the platform, i.e., $\varepsilon \in [0,1]$.} 

If any publisher's NFT is referenced by another publisher's NFT, the interest rate growth is delayed with a probability of $\varepsilon$, providing an additional income source for the NFT owner.
This allows any publisher to adjust their payment $\pi_{r}$, thereby affecting both the decay parameter $d$ and the interest rate $q$.
Therefore, the action for any publisher who is about to publish their NFT $\mathcal{R}_{i,h,\Theta}$ becomes a tuple of three elements, represented as
\begin{equation}
\mathcal{A}_{\text{out}}=(\pi_r, w_1, w_2, \dots, w_{|\vec{\Theta}|}), \text{ with}\sum_{i=1}^{|\vec{\Theta}|} w_i=1-w_0,w_{i,\forall i} \in \mathbb{R}_0^+.
\label{eq:a_out}
\end{equation}


Furthermore, when a publisher publishes an NFT, the action of setting the reference ratio for each component can reflect another characteristic of the outcome action $\mathcal{A}_{\text{out}}$. This is mapped to:
\begin{equation*}
\Phi(w_i)=\prod_{k=i}^{|\vec{\Theta}|} w_k \varpropto \varepsilon, \forall i > 0,
\end{equation*}
\textit{where $\Phi(w_i)$ exclusively serves as the income's coefficient but also correlates to the probability $\varepsilon$ of the NFT being referred each time when a new NFT is published in the network.} As a result, the equation produces an income coefficient $\varepsilon$. In practice, we can regard the income coefficient $\varepsilon$ as the \textit{quality} of an NFT.

\subsection{Income Function}\label{subsec-income}
The position of an NFT on the blockchain determines its income derived from subsequent rounds. The set of freshly minted NFTs at round $h+1$, denoted as $\mathcal{I}_{(h+1)}$, refers to $\mathcal{R}_{i,h,\Theta}$.

The income action $\mathcal{A}_\text{in}$ is represented by the unit price being set for a subsequent one that references the current $\mathcal{R}_{i,h,\Theta}$:

\begin{equation*}
\mathcal{A}_\text{in}=\psi_{\mathcal{R}},
\end{equation*}
\textit{where $\psi_{\mathcal{R}} \in (0, \bar{\psi_{\mathcal{R}}}$] with $\bar{\psi_{\mathcal{R}}}$ being the maximum value of the unit price, defined as the cost for using a minimal fraction of a resource, e.g., 1\% of the resource.}

The revenue for round $h+1$ is determined by $k\sigma^{-1}|\mathcal{I}_{(h+1)}|$, where $k$ is a constant. The income function now incorporates the income coefficient $\varepsilon$ as follows:
\begin{equation*}
\small
\begin{aligned}
\mathcal{I}_{\mathcal{R}_{i,h,\Theta}}
= & k\sigma^{-1} |\mathcal{I}_{(h+1)}|\varepsilon + k\sigma^{-2} |\mathcal{I}_{(h+2)}|\varepsilon + ... + k\sigma^{-d} |\mathcal{I}_{(h+d)}|\varepsilon + \hat{\mathcal{I}}\\
=& \varepsilon\sum_{j=1}^{d} k l(\hat{\sigma}, \pi_r)^{-j} |\mathcal{I}_{(h+j)}|+ \hat{\mathcal{I}},\\
\end{aligned}
\end{equation*}
\textit{where $\hat{\mathcal{I}}>0$ is an initial fixed reward if the item represented by the new NFT $\mathcal{R}_{i,h,\Theta}$ is more performant than that of any of the items contained in the candidate set $\Omega_{\mathcal{R}_{i,h,\Theta}}$.}

\subsection{Back to Payoff Function}\label{subsec-payoff}

The publisher's payoff function $\mathbb{U}$ now takes into account the unit prices set by the publisher and their impact on the income coefficient, calculated as follows:

\begin{equation*}
\footnotesize
\begin{aligned}
\mathbb{U}_{\mathcal{R}_{i,h,\Theta}}
= & \varepsilon\sum_{j=1}^{d} k\sigma^{-j}|\mathcal{I}_{(h+j)}| + \hat{\mathcal{I}} - \lambda p_0 - \pi_r-\sum_{j=1}^{d}(1+q)^{j-\varepsilon} \frac{p_0 (1-\lambda)}{d} \\
= & \varepsilon\sum_{j=1}^{d} k\sigma^{-j}|\mathcal{I}_{(h+j)}| + \hat{\mathcal{I}} - p_0\left[ \lambda+ \sum_{j=1}^{d}(1+q)^{j-\varepsilon} \frac{p_0 (1-\lambda)}{d} \right]-\pi_r \\
& \propto d \lbrack \sigma^{-d} |\mathcal{I}| - (1+q)^{d}\rbrack.
\end{aligned}
\end{equation*}

As a result, the optimization pricing problem can be reformulated over many rounds, where each round corresponds to a publisher releasing a new NFT $\mathcal{R}_{i,h,\Theta}$. This setup is illustrated as follows:


\begin{equation}
\begin{aligned}
&\max_{\mathcal{A}_{\text{in}},\mathcal{A}_{\text{out}}} \quad
\mathbb{U} 
= \sum_{\mathcal{R}_{i,h,\Theta}}\left[\varepsilon\sum_{j=1}^{d} k\sigma^{-j}|\mathcal{I}_{(h+j)}|+ \hat{\mathcal{I}} \right. \\
&\left. - p_0\left[ \lambda+ \sum_{j=1}^{d}(1+q)^{j-\varepsilon} \frac{(1-\lambda)}{d} \right]-\pi_r\right]\\
    s.t. \quad & (a): \sigma=l(\hat{\sigma},\pi_r), (1+q)=g(\hat{q},\pi_r), d=\hat{d}+f(\pi_r),\\
    & (b): \hat{d} = \phi(1-\lambda),\\
    & (c): 0\leq \lambda,\sigma,\varepsilon \leq 1,\\
    & (d): d\in \mathbb N,\\
    & (e): 0 \leq \pi_r \leq \bar{\pi_r}.
\end{aligned}
\label{eq:optimization}
\end{equation}

The twofold benefits of this approach are clear. First, the publisher invests more to generate a higher income due to the enhanced performance of the resources that others obtain, coupled with the delay in the growth of the outcome interest and the income descending. Second, if the publisher is confident about the performance of their resources, they can choose to set a higher unit price, resulting in increased profits during the income phase. 


\subsection{Game Analysis}
\label{sec-ananlysis}

We then provide theorems ensuring that the distribution process is completed within a finite number of rounds. We consider the game to be a non-convex and NP-hard problem, and Nash Equilibrium (NE) exists in the considered game.

\begin{thm}\label{lem-distri}
(Finality) If $\sigma \in (0,1)$ and $d \in \mathbb{N}$, the distribution of $\mathcal{I}_{\mathcal{R}_{i,h,\Theta}}$ concludes within $d$ rounds.
\end{thm}

\begin{prf}[\textit{Theorem~\ref{lem-distri}}]
Given $|\sigma|<1$, the sum of a geometric series $\sum_{i=0}^{d} \sigma^{i}$ converges to $\frac{1-\sigma^{d+1}}{1-\sigma}$. Therefore, as $\sigma \in (0,1)$ and $d \in \mathbb{N}$, the sum of the series is a finite value, ensuring the distribution of $\mathcal{I}_{\mathcal{R}_{i,h,\Theta}}$ ends within $d$ rounds. \qed
\end{prf}

\textit{Theorem~\ref{lem-distri}} implies that revenue derived from each minted NFT within this network is positively correlated to the total participants ($\sum_d|\mathcal{I}|$) or valid references ($|\Theta|$), and inversely related to the round interest $q$. For instance, when an NFT is procured at a higher initial price ($\lambda$), fewer rounds are needed to compensate the preceding NFTs. Conversely, a higher descending rate ($\sigma$) results in a reduced number of benefit-yielding rounds ($d$).

However, the comprehensive payoff function, encapsulating all parameters, including $|\mathcal{I}|$, $\lambda$, $d$, $w$, and $\sigma$, is inherently complex. Foreseeing the dominant parameter becomes increasingly difficult as a sudden influx of participants ($|\mathcal{I}|$) could offset the influence of other parameters. This abrupt increase could be due to external stimuli, which are generally not known from the system's perspective. Moreover, the selection of $\sigma$ carries significant weight for the platform's stability. A larger $\sigma$ distributes more income to earlier rounds. Thus, it is imperative to balance between platform growth and publisher engagement, leading to an optimization problem to maximize publisher engagement while ensuring the platform's healthy development.

\begin{thm}\label{lem-convex}
(Complexity) The optimization of the payoff function $\mathbb{U}$ for any NFT minted is a non-convex, non-strict NP-hard problem.
\end{thm}

\begin{prf}[\textit{Theorem~\ref{lem-convex}}]
The proof, following the principles discussed in a previous study~\cite{wang2023referable}, involves examining the Hessian matrix of $\mathbb{U}$ with respect to $\sigma$ and $q$. We denote $A=\frac{\partial^2 \mathbb{U}}{\partial \sigma^2}$, $B=\frac{\partial^2 \mathbb{U}}{\partial \sigma \partial q}$, and $C=\frac{\partial^2 \mathbb{U}}{\partial q^2}$. The condition for convexity requires the Hessian matrix to be positive semi-definite, which in turn demands that all its eigenvalues be non-negative. This leads to the condition $AC-B^2 \geq 0$. However, we find that $AC-B^2 \ngeq 0$ for $\sigma$ and $q$ in the range [0,1]. This indicates the non-convexity of the optimization problem. In this case, we categorize the optimization as non-strict NP-hard, acknowledging the complexity and potential computational challenges in finding optimal solutions. Hence, traditional convex optimization methods are not feasible for optimizing $\mathbb{U}$.
\qed
\end{prf}

The implications of \textit{Theorem~\ref{lem-convex}} are profound in understanding the complexity of the payoff function $\mathbb{U}$, where the complexity arises due to its dependence on numerous parameters such as the number of participants $|\mathcal{I}|$, initial price $\lambda$, round number $d$, weight $w$, and descending rate $\sigma$. Moreover, the nature of this problem makes it challenging to identify a dominant parameter, especially in scenarios characterized by a sudden surge in participants.

\begin{thm}\label{lemma-stack}
(Solvability) The considered NFT pricing game can be transformed from a stochastic optimization problem into a repeated $N$-player non-cooperative game~\cite{fujiwara2015non} with finite actions $a_{\mathcal{R}_{i,h,\Theta}}$ that satisfies Mixed Strategy Nash Equilibrium
(MNE)~\cite{reny1999existence}.
\end{thm}

\begin{prf}[\textit{Theorem~\ref{lemma-stack}}]
As proved in \textit{Nash's theorem}~\cite{nash-1}, a finite non-cooperative game has at least one NE solution for agents taking mixed strategies. This is expressed as:

\begin{equation*}
    \Delta(\mathcal{A}_j)=
    \left\{
    p_j = \lbrace p_{j,1},\dots,p_{j,k}\rbrace,p_{j,k}\geq 0,\sum_kp_{j,k}=1
    \right\},
\end{equation*}
\textit{where $\Delta(\mathcal{A}_j)$ denotes the probability distribution of publisher $j$'s action space $\mathcal{A}_j$ for all possible actions, and $p_{j,k}$ is the probability of the $j$-th publisher taking the $k$-th action.}
This implies that an MNE point can be eventually found in the considered pricing game. Let $a_{\mathcal{R}_{i,h,\Theta}}^{j,\star}$ denote the optimal outcome and income actions of publisher $j$ regarding her NFT published on $h$-th round with a ratio list of $\Theta$. The point $a_{\mathcal{R}_{i,h,\Theta}}^{j,\star}$ can be considered as an MNE if it satisfies:

\begin{equation*}
\mathbb{U}^j_{\mathcal{R}_{i,h,\Theta}}(a_{\mathcal{R}_{i,h,\Theta}}^{j,\star}) \geq \mathbb{U}^j_{\mathcal{R}_{i,h,\Theta}}(a^j_{\mathcal{R}_{i,h,\Theta}}), \forall i,j,h,\Theta.
\label{eq:mne}
\end{equation*}\qed
\end{prf}

Reaping profits while setting prices against unknown competitors and solving the optimization problem are challenging for any publisher $j$. Any adjustments of its strategy of $\mathcal{A}_{\text{in}}$ and $\mathcal{A}_{\text{out}}$ regarding an NFT $\mathcal{R}^j_{i,h,\Theta}$ are intricately coupled with an unknown number of the strategies applied by different publisher $j'$. It is a dynamic pricing environment in which any publisher $j$ needs to solve the payoff function and work out the most profitable pricing strategy independently.

To this end, we propose that each publisher can implement \textit{Theorem~\ref{lemma-stack}} by running a decentralized DRL algorithm to play this non-cooperative game, as outlined in the following section.

\section{Approaching Optimal Utility via DRL}
\label{sec-drl}

While achieving equilibrium as described in \textit{Theorem~\ref{lemma-stack}} entails high computational complexity (NP-hard), practical methods can still approach an approximate solution (stably converged to a value, which we consider as \textit{practically solvable} in this study).

\smallskip
\noindent\textbf{DRL approach.} Model-free reinforcement learning (RL) offers a dynamic solution for effective machine learning (ML)-based decision-making, capable of tackling intricate optimization problems in changing environments, including the pricing problem discussed in Sec.\ref{sec-ananlysis}. DRL extends this capability, leveraging deep neural networks to manage intricate data within large state-action spaces and to execute complex function approximation tasks effectively. This makes it particularly proficient in handling the decentralized Stackelberg problem, a type of sequential decision-making challenge. To operationalize \textit{Theorem~\ref{lemma-stack}}, the preferred approach is policy-based deep reinforcement learning (DRL) algorithms.

\subsection{DRL Model}
\label{subsec-drl-model}

We present the necessary components in such a DRL model:

\smallskip
\noindent\textbf{Agents.} The entities that execute a learning process and maintain a local prediction model are the agents. We suppose each publisher $j$ who participates in the NFT network or has performed some actions is an agent in this paper.

\smallskip
\noindent\textbf{Episode.} The determination of a concluded episode hinges on whether the learning progression employs a Monte-Carlo update or a Temporal-Difference update, as referenced in the literature~\cite{rl-survey}. In the context of this paper, the pricing contest within the network is depicted as a recurrent game with no predefined termination point. During an episode, a complete cycle takes place in which a publisher prepares to release their NFT after examining the current state. As soon as the NFT gets published, the publisher determines the subsequent actions based on the new state.  The interval between identifying two successive states can be set at a specific duration, e.g., the full cycle of a block's height that stores NFTs.

\smallskip
\noindent\textbf{States ($\mathcal{S}$).} Each agent (or publisher) obtains the continuous observation of the environment, denoted by $s \in \mathcal{S}$. 
Each publisher maintains a current state in a given episode, specifically in relation to the NFT that is about to be sent, as given by $s_{\mathcal{R}_{i,h,\Theta}}$. 
The set of allowable states for each impending NFT release is closely associated with a predetermined NFT candidate set $\Omega_{\mathcal{R}_{i,h,\Theta}}$. 
This candidate set is defined by a specific quantity of NFTs that have not yet expired and are sampled according to the quality $\varepsilon$ of $\mathcal{R}_{i,h,\Theta}$.
This state consists of two elements:
\begin{itemize}
    \item the unit price of the candidate NFTs previously published\footnote{We consider that ``one gets what one pays for'' is applied during the sharing. This indicates that the proportional mapping correlation between the unit price and the resource quality can be established.}. $[\psi]_{i',h',\Theta}$ where $i' \neq i$ and $h' < h$; and

    \item the resource demands from the previous observation of other NFTs $[\pi_r, w]_{i',h',\Theta}$ where $i' \neq i$ and $h' < h$.
\end{itemize}
These two components together define the publisher's valid state set, as given by
\begin{equation*}\label{eq:action-space}
    s_{\mathcal{R}_{i,h,\Theta}}=\lbrace \psi, [\pi_r, w]\rbrace_{i',h',\Theta}.
\end{equation*}

\smallskip
\noindent\textbf{Policy $\pi(\cdot)$.} The local policy is persistently trained and updated over time. In this process, each NFT publisher supplies their local learning model with his/her observed state $s$ from the environment across various episodes. Consequently, a probability distribution over valid actions by any publisher can be derived, which aids in making decisions in subsequent episodes. This relationship is represented as $\pi(s,a): \mathcal{S} \to \mathcal{A}$.

\smallskip
\noindent\textbf{Actions ($\mathcal{A}$).} 
In reference to the action definition in the non-cooperative game involving $N$ agents, every agent forms its decision $a\in \mathcal{A}$ based on its local policy. This policy is generated by the agent's local learning model during the ongoing episode. The set of permissible actions for publisher $j$ has three components:
\begin{itemize}
    \item the decision $\mathcal{A}_{\text{trigger}}$ of triggering to publish an NFT or not in the current episode;

    \item the outcome action, which includes the sum of payment designed to adjust the interest and descending rate favorably for the publisher, as well as the segments of a previous NFT that his NFT is set to reference;

    \item the income action, which is the unit price specified for his NFT in anticipation of it being referred by others.
\end{itemize}
These three components together define the publisher's valid action set, as given by
\begin{equation*}\label{eq:action-space}
\begin{aligned}
    a_{\mathcal{R}_{i,h,\Theta}}=\lbrace
    & \mathcal{A}_{\text{trigger}}, \mathcal{A}_{\text{down-pay}},\mathcal{A}_{\text{out}},\mathcal{A}_{\text{in}}\rbrace,
\end{aligned}
\end{equation*}
\textit{where $\mathcal{A}_{\text{trigger}}$ is a Boolean value; 
$\mathcal{A}_{\text{down-pay}}$ reflects the down-payment ratio $\lambda \in \lbrack 0,1\rbrack$.}


\smallskip
\noindent\textbf{Reward function ($\mathsf{R}$).} A crafted reward function provides valuable insights about the mapping $\mathcal{S} \to \mathcal{A}$, thereby enhancing the effectiveness of the local policy. Taking into consideration the optimization problem outlined in~(\ref{eq:optimization}), any NFT publisher has two primary goals in an episode: to minimize the cost of publishing an NFT related to its resource, and to maximize the income from the potential usage of this resource by others in the network. Hence, the reward function is constructed as
\begin{equation*}\label{eq:reward-function}
    \mathsf{R}(a_{\mathcal{R}_{i,h,\Theta}}) = \mathbb{U}_{\mathcal{R}_{i,h,\Theta}}.
\end{equation*}
All NFT publishers are considered \textit{rational}. Their objective is to maximize their rewards throughout the process of resource trading.

\smallskip
\noindent\textbf{Environment.} An environment is viewed as a black box that takes as input the \textit{actions} of the \textit{agents} and outputs the \textit{states}. In our paper, the NFT network is the environment shared by all miners (in PoW contexts) or validators (PoS).

\subsection{Operation Sketch}
\label{subsec-drl-solve}

\begin{algorithm*}[!hbt]
\scriptsize
\caption{Implementing \textit{\textbf{the (DRL-based) reference incentive}} model}\label{algo-1}
\vspace{-0.2in}
\SetAlgoLined
\begin{multicols}{2}[\raggedcolumns]
\normalem 
\textbf{Initialization} 

Initialize DRL parameters; Reset DRL replay memory; Clean caches of $(\mathcal{O}_{\mathcal{R}_{i,h,\Theta}}, \mathcal{I}_{\mathcal{R}_{i,h,\Theta}})$.

\SetKwFunction{MainFunc}{\textcolor{teal}{$\mathsf{MainFunc}$}}
\SetKwFunction{RetrieveResources}{\textcolor{teal}{$\mathsf{RetrieveResources}$}}
\SetKwFunction{MemorizeExperience}{\textcolor{teal}{$\mathsf{MemorizeExperience}$}}
\SetKwFunction{SampleExperience}{\textcolor{teal}{$\mathsf{SampleExperience}$}}
\SetKwFunction{ObserveNetwork}{\textcolor{teal}{$\mathsf{ObserveNetwork}$}}

\SetKwProg{firstproc}{Procedure}{}{}
\firstproc{\MainFunc{}}{
\While{True}{
    \ForEach{publisher $j$ in the network, \textbf{\textit{concurrently}}}{
        $\Omega_{\mathcal{R}_{\text{raw}}}$ $\gets$ \RetrieveResources{}, $\forall j $ omitted\;
    
        $s_{\mathcal{R}_{\text{raw}}}$ $\gets$ $\Omega_{\mathcal{R}_{\text{raw}}}$;
        
        $a_{\mathcal{R}_{\Theta}}$ $\gets$ $\text{DRLAgent.ChooseAction}(s_{\mathcal{R}_{\text{raw}}})$;
    
        $(\textit{TX}_{id}, s_{\mathcal{R}_{i,h,\Theta}}, s'_{\mathcal{R}_{i,h,\Theta}}, a_{\mathcal{R}_{i,h,\Theta}}, \mathcal{O}_{\mathcal{R}_{i,h,\Theta}}, \mathcal{I}_{\mathcal{R}_{i,h,\Theta}})$ $\gets$ $\text{DRLAgent.Execute}(a_{\mathcal{R}_{\Theta}})$, where $\text{Quality}_i$ and $\text{Price}_i$ $\in$ $a_{\mathcal{R}_{i,h,\Theta}}$;

        \If{$a_{\mathcal{R}_{i,h,\Theta}}.\mathcal{A}_{\text{trigger}}$ is $False$}{
            \MemorizeExperience{}, $\forall j $ omitted\;
        } \Else{
            \While{current block height $< (h+d)$, \textbf{\textit{non-blockingly}}}{
                $\mathcal{O}_{\mathcal{R}_{i,h,\Theta}}$.Update($q,d$) \textbf{and} $\mathcal{I}_{\mathcal{R}_{i,h,\Theta}}$.Update($\sigma,d$);
            }
            \If{current block height $\geq (h+d)$}{
                \MemorizeExperience{}, $\forall j $ omitted\;
            }
        }
    
        \text{DRLAgent.UpdateParameters}(\SampleExperience{}), $\forall j $ omitted\;

        \ObserveNetwork{}, $\forall j $ omitted\;
    }
}

}\textbf{end}

\SetKwProg{secondproc}{Procedure}{}{}
\secondproc{\RetrieveResources{}}{
  \While{$|\Omega_{\mathcal{R}_{\text{raw}}}| < |\Omega_{\textit{MAX}}|$}{
    $\Omega_{\mathcal{R}_{\text{raw}}}.$append(($\text{Quality}_{i'}$, $\text{Price}_{i'}$)), $\forall i'$-th NFTs in a sliding window that adjusts with the growth of the blockchain tip;
  }
}\textbf{end}

\SetKwProg{thirdproc}{Procedure}{}{}
\thirdproc{\MemorizeExperience{}}{
        $\text{DRLAgent.Memorize}((\textit{TX}_{id}, s_{\mathcal{R}_{i,h,\Theta}}, s'_{\mathcal{R}_{i,h,\Theta}}, a_{\mathcal{R}_{i,h,\Theta}}, \mathcal{O}_{\mathcal{R}_{i,h,\Theta}}, \mathcal{I}_{\mathcal{R}_{i,h,\Theta}}))$;
}\textbf{end}

\SetKwProg{fourthproc}{Procedure}{}{}
\fourthproc{\SampleExperience{}}{
        $(\textit{TX}_{id}, s_{\mathcal{R}_{i,h,\Theta}}, s'_{\mathcal{R}_{i,h,\Theta}}, a_{\mathcal{R}_{i,h,\Theta}}, \mathcal{O}_{\mathcal{R}_{i,h,\Theta}}, \mathcal{I}_{\mathcal{R}_{i,h,\Theta}}) \gets \text{DRLAgent.Sample}(\textit{BATCH\_SIZE})$;

        $\mathsf{R}(a_{\mathcal{R}_{i,h,\Theta}}) \gets \mathcal{I}_{\mathcal{R}_{i,h,\Theta}} - \mathcal{O}_{\mathcal{R}_{i,h,\Theta}}$, with $\mathsf{R}(a_{\mathcal{R}_{i,h,\Theta}}) = \mathbb{U}_{\mathcal{R}_{i,h,\Theta}}$;


        \textbf{return} $(s_{\mathcal{R}_{i,h,\Theta}}, s'_{\mathcal{R}_{i,h,\Theta}}, a_{\mathcal{R}_{i,h,\Theta}}, \mathsf{R}(a_{\mathcal{R}_{i,h,\Theta}}))$;
}\textbf{end}

\SetKwProg{fifthproc}{Procedure}{}{}
\fifthproc{\ObserveNetwork{}}{
    \While{blockchain is growing}{
        Observe the network to identify interesting NFTs, publish own NFTs based on these findings, and trigger the next iteration inside \MainFunc{} for each publisher $j$.
    }
}\textbf{end}

\textcolor{gray}{Notice: ``DRLAgent.Execute($\cdot$)'' in the context of our reference incentive model, represents the action where publisher $ j $ releases their new rNFT $ \mathcal{R}_{i,h,\Theta} $ to the blockchain. This includes the available rNFT index $ i $, block height index $ h $, and reference list $ \Theta $.}

\end{multicols}
\vspace{-0.15in}
\end{algorithm*}

Algorithm \ref{algo-1} operates autonomously for each publisher, ensuring the system's effective convergence to a stationary point even in a fully decentralized setup. Each publisher begins by setting up their experience replay memory, initializing predefined hyperparameters, weights, and any cumulative counters associated with an NFT (Line 2). At the outset, publisher $j$ selects an action $a$ at random. During the commencement of each workflow, the publisher scans the network for appealing NFTs within a sliding window linked to the blockchain's tip, assessing the data quality and pricing of these NFTs (Line 6). When a decision is made to publish an NFT, i.e., $\mathcal{A}_{\text{trigger}}$ is set to true, the publisher also determines the down-pay ratio ($\mathcal{A}_{\text{down-pay}}$), references to preceding NFTs ($\mathcal{A}_{\text{out}}$), and sets the price for the new NFT ($\mathcal{A}_{\text{in}}$) (Line 9). 
Note that the references to preceding NFTs undergo a normalization process to ensure that the total payment ratio distributed among the referred NFTs sums up to one, in accordance with Equation (\ref{eq:a_out}).
The cumulative outcome and income for this NFT are updated along with the reference payment over time (Line 15). The reward for this new NFT might be immediate or might require waiting until the decay parameter $d$ concludes (Lines 10--20), depending on the publisher's decision. This event is then recorded in the replay memory. Consequently, each publisher's DRL agent updates its parameters by sampling from the memory. This training cycle continues indefinitely or until a specified maximum number of block heights is reached.

\smallskip
\noindent\textbf{Algorithm's complexity}. The computational complexity of this model can be expressed as $O((N+1) \times (101 \times 10^k))$, where $N$ represents the number of NFTs to be referenced, and $k$ signifies the decimal precision. This complexity accounts for scenarios where the number of NFTs ranges from 0 to $N$, each NFT possesses a reference rate varying from 0\% to 100\%. This intricate calculation poses significant challenges, particularly in contexts demanding continuous states and actions, especially when computational resources are limited. To this end, it becomes imperative to constrain $N$ within practical bounds while maintaining a high level of accuracy denoted by $k$. This is crucial for the effective deployment of the model in real-world applications where resource constraints are prevalent. To illustrate the efficacy of our approach within these constraints, we focus our attention on a specific use case: limiting the number of references from an unlimited set to just two. This targeted scenario not only showcases the practicality of our approach but also provides a tangible example of its effectiveness. We present experimental results in the following part to practically substantiate the utility and performance of our proposed DRL method.
\section{Evaluation}\label{sec-experiment}

In this section, we experimentally assess the DRL-based solutions by applying our incentive model to a use case. We examine convergence performance and stability under a variety of different environment settings. Several insights are also shed and discussed.

\subsection{Experiment Design}

\begin{figure}[t]
    \centering
    \includegraphics[width=1\linewidth]{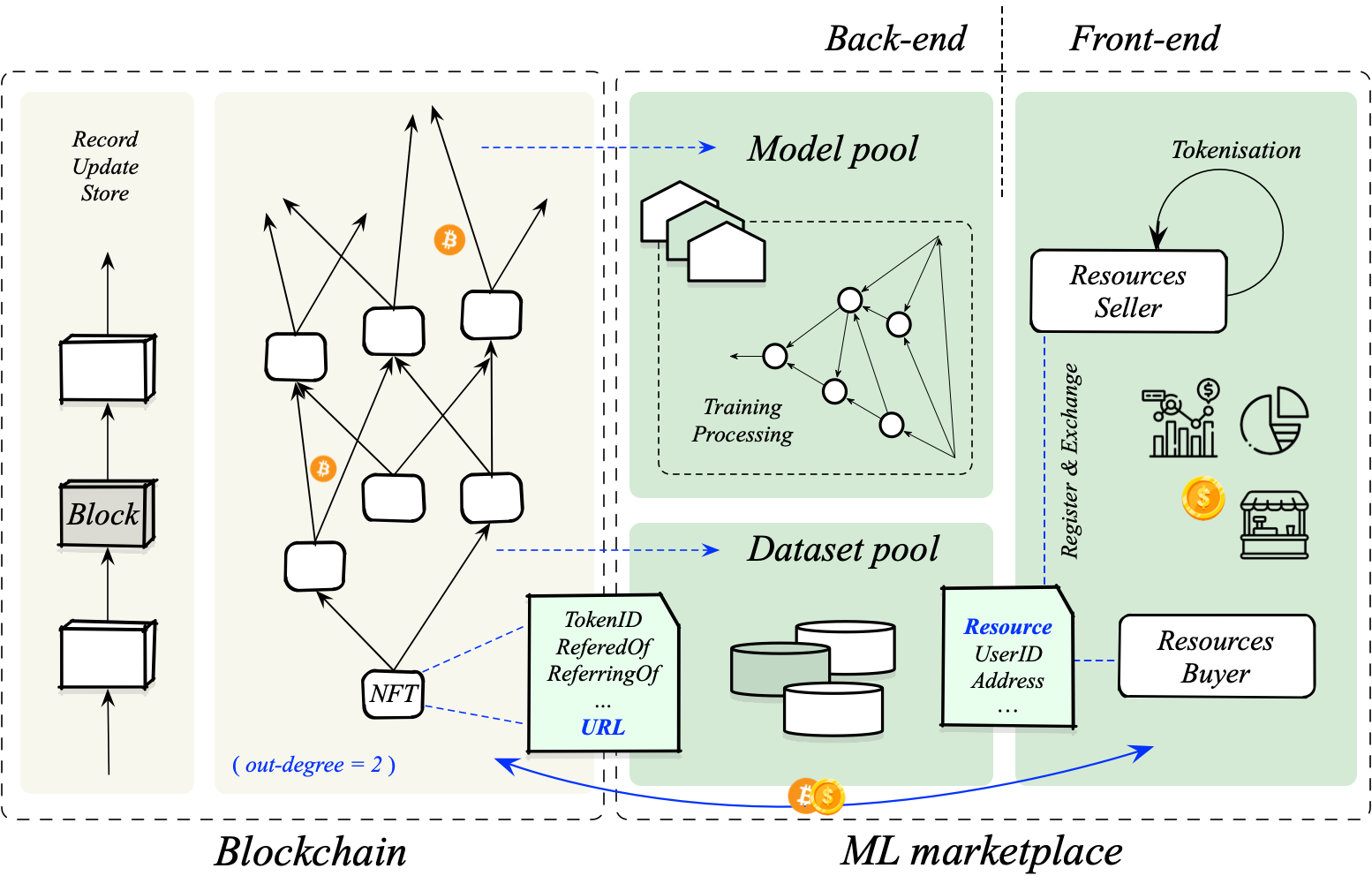}
    \vspace{-0.25in}
    \caption{Use Case: An ML Resource Marketplace}
\label{fig:sys_arch_dsp}
\vspace{-0.2in}
\end{figure}

In our reference incentive model, we specifically focus on references between NFTs representing \textit{datasets} and \textit{models} within an ML resource marketplace (see Fig.\ref{fig:sys_arch_dsp}). To address the complexity of multiple datasets/models, a pre-merging mechanism can be applied to align with the EIP-\hlhref{https://eips.ethereum.org/EIPS/eip-998}{998} standard for composability. The mechanism allows an NFT to represent a composite of multiple datasets and/or models. Thus, each NFT can encapsulate a variety of dataset and model resources as a single entity for reference\footnote{This enables the recursive finalization of profit distribution by sequentially decapsulating bundled resources. The specifics of inner profit distribution, however, fall outside the scope of our study, as they can be implemented by Apps behind the scenes.}. This mechanism is especially beneficial for Federated Learning (FL) scenarios. It facilitates a streamlined approach by representing datasets and models as a single composable NFT entity and allowing underlying applications to deconstruct and process the bundled resources efficiently for critical FL operations such as Federated Averaging.

\smallskip
\noindent\textbf{Parameter settings.} Our experiments include a series of parameters: the number of publishers $N$, the quality distribution of the initial NFTs, the initial raw interest $\hat{q}$, the size of the candidate set $\Omega_{\mathcal{R}_{i,h,\Theta}}$ of any NFT $\mathcal{R}_{i,h,\Theta}$, the initial decay parameter $d$, the initial fixed reward $\hat{\mathcal{I}}$, and the initial fixed expense $\hat{\mathcal{O}}$.

For the purpose of proof-of-concept, $p_0$ is converted to:
\begin{equation*}
\begin{aligned}
p_0
= & \left(\hat{\mathcal{O}} + \mathcal{O}\right)\times\lbrack w_0, w_d, w_m\rbrack, \text{ 
with }\mathcal{O}\propto \lbrace w_d,w_m\rbrace,
\end{aligned} 
\end{equation*}
\textit{where $w_d$ and $w_m$ are the reference weights for a dataset and a model, respectively}. Note that the term ``bundle'' is omitted for simplicity in the presentation.

The outcome action is accordingly converted to:
\begin{equation*}
\mathcal{A}_{\text{out}}=(\pi_r, w_d, w_m), \text{ with }w_0+w_d+w_m=1,\forall w \in \mathbb{R}_0^+.
\end{equation*}
\textit{where $w_d$ and $w_m$ correspond to a portion of a dataset and a model that is being shared, respectively.}



\smallskip
\noindent\textbf{Assumptions.} In the context of the ML resource marketplace where our incentive reference model is applied, we assume that each user knows which is most suitable for the whole platform, and each NFT uses at most one dataset and one model as its basis. A portion of a model might be pruned or compressed, and a dataset might be a partial dataset. For example, 50\% of a dataset might represent half the data size, and 50\% of a model might mean pruning half of the parameters. Both the dataset and model resources are formatted as 256-bit hash values, serving as URIs that redirect to decentralized data storage platforms such as IPFS.

\smallskip
\noindent\textbf{Hardware settings.}
Experiments are conducted on nodes in a high-performance computing server with the following specifications:

\begin{itemize}
    \item CPU. 2 x Intel(R) 6138 CPU @ 2.00GHz, 2 × 40 cores
    \item GPU. 8 × NVIDIA PCIe A100
    \item RAM. 250GB
    \item Disk. 2TB
\end{itemize}

\smallskip
\noindent\textbf{Software settings.}
\label{subsubsec: software}
We carry out the experiments upon the Ubuntu 20.04 OS environment with PyTorch 2.0.1 in Python 3.10.13.



We have a predefined number of agents operating concurrently on nodes, with a decentralized pricing game running in the same cluster for simplicity and generality. The solution employs a continuous version of Proximal Policy Optimization (PPO), known for its effectiveness in DRL. It is important to note that our solution is adaptable; it works seamlessly with various DRL algorithms, including the classical Deep Deterministic Policy Gradient (DDPG), ensuring transparency and flexibility in implementation.

\subsection{Experiment Results}
\begin{figure*}[t]
    \centering
    \subfigure[$N=10$]{
    \begin{minipage}[t]{0.23\textwidth}
    \centering
    \includegraphics[width=1.7in]{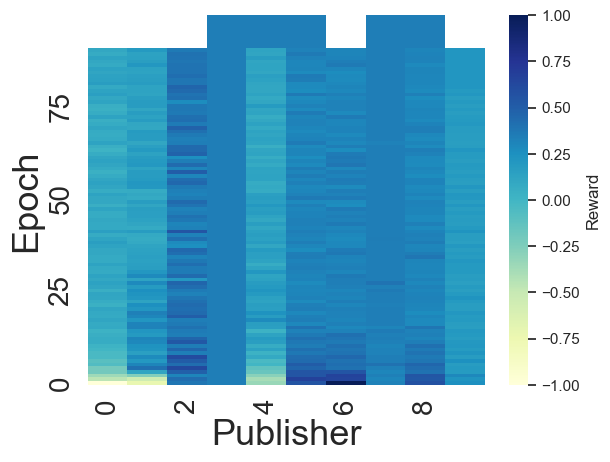}
    \end{minipage}
    \label{fig:publisher_num_10}
    }
    \subfigure[$N=20$]{
    \begin{minipage}[t]{0.23\textwidth}
    \centering
    \includegraphics[width=1.7in]{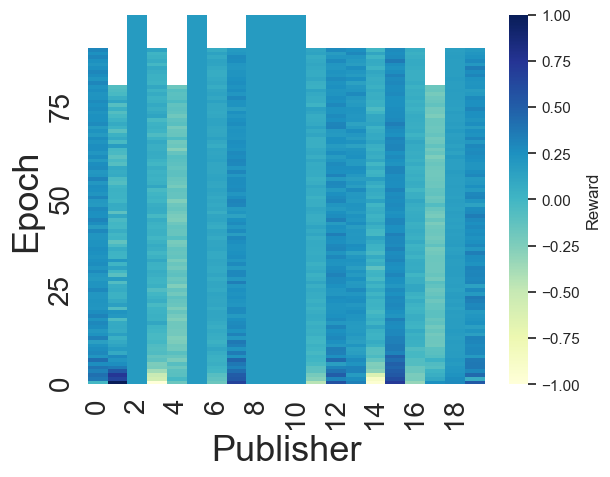}
    \end{minipage}
    \label{fig:publisher_num_20}
    }
    \subfigure[$N=30$]{
    \begin{minipage}[t]{0.23\textwidth}
    \centering
    \includegraphics[width=1.7in]{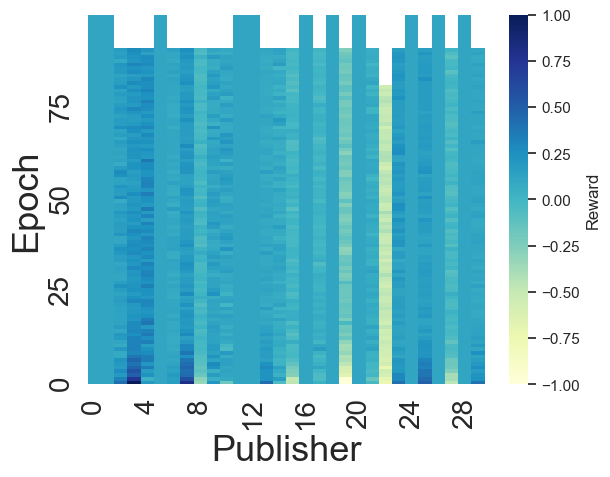}
    \end{minipage}
    \label{fig:publisher_num_30}
    }
    \subfigure[$N=40$]{
    \begin{minipage}[t]{0.23\textwidth}
    \centering
    \includegraphics[width=1.7in]{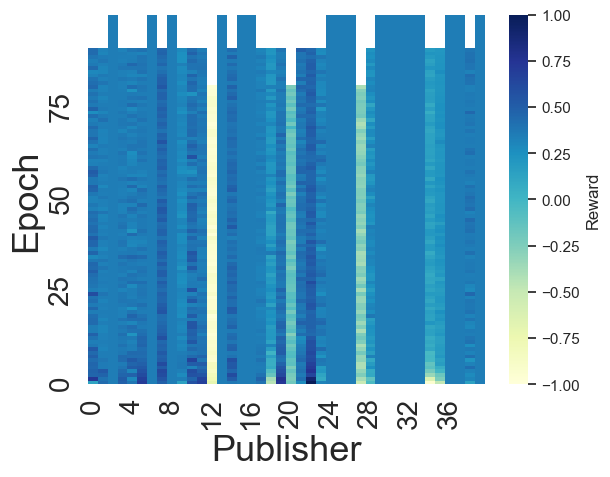}
    \end{minipage}
    \label{fig:publisher_num_40}
    }
        \vspace{-0.2in}
\caption{Comparison between different settings of the number of publishers $N$ in terms of normalized reward with a setting of uniform-distributed initial NFTs, $\hat{q}=0.01$, $|\Omega_{\mathcal{R}_{i,h,\Theta}}|=10$, $\hat{d}=10$, $\hat{\mathcal{I}}=2$, $\hat{\mathcal{O}}=0.1$.}
\label{fig:publisher_num}
\end{figure*}

\begin{figure*}[t]
    \centering
    \subfigure[Uniform Distribution]{
    \begin{minipage}[t]{0.23\textwidth}
    \centering
    \includegraphics[width=1.7in]{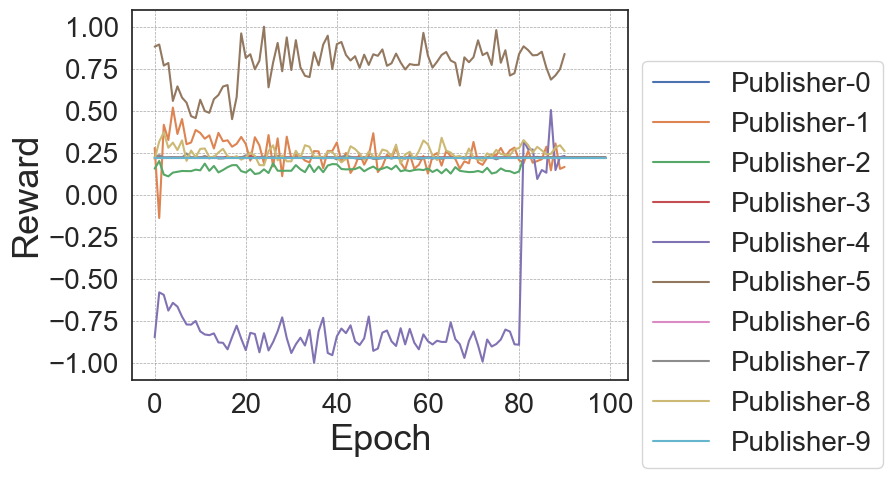}
    \end{minipage}
    \label{fig:uniform_distribution}
    }
    \subfigure[Normal Distribution]{
    \begin{minipage}[t]{0.23\textwidth}
    \centering
    \includegraphics[width=1.7in]{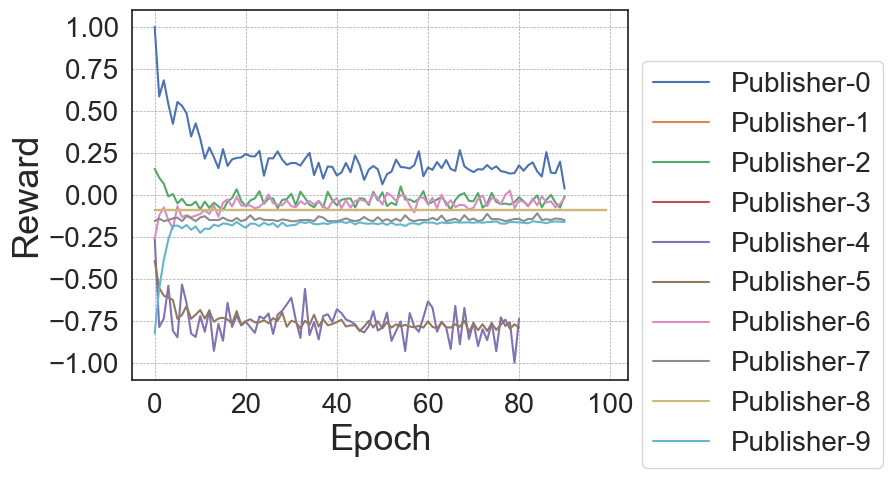}
    \end{minipage}
    \label{fig:normal_distribution}
    }
    \subfigure[Pareto Distribution]{
    \begin{minipage}[t]{0.23\textwidth}
    \centering
    \includegraphics[width=1.7in]{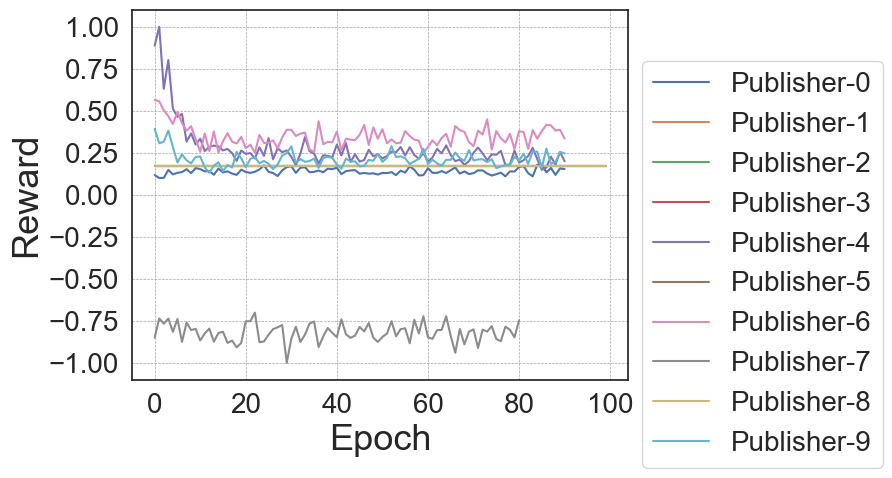}
    \end{minipage}
    \label{fig:pareto_distribution}
    }
    \subfigure[Poisson Distribution]{
    \begin{minipage}[t]{0.23\textwidth}
    \centering
    \includegraphics[width=1.7in]{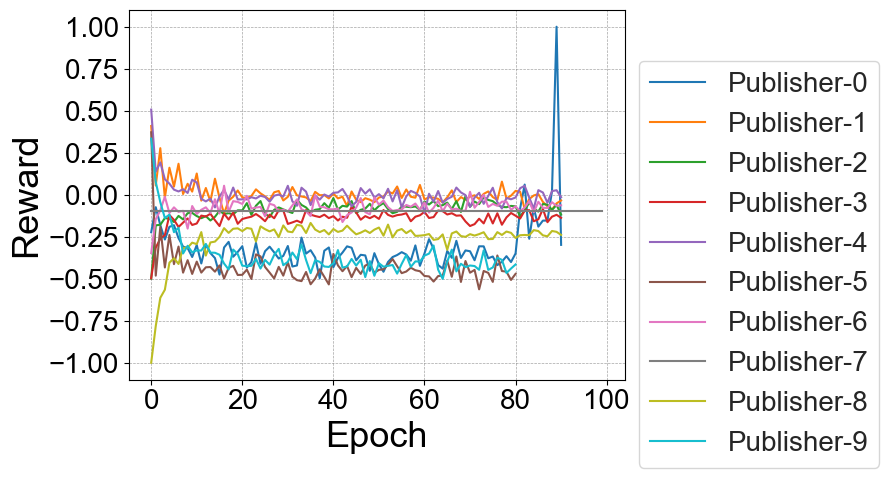}
    \end{minipage}
    \label{fig:poisson_distribution}
    }
    \vspace{-0.2in}
\caption{Comparison between different settings of the \textcolor{teal}{quality distribution of initial NFTs} in terms of normalized reward with a setting of $N=10$, $\hat{q}=0.01$, $|\Omega_{\mathcal{R}_{i,h,\Theta}}|=10$, $\hat{d}=10$, $\hat{\mathcal{I}}=2$, $\hat{\mathcal{O}}=0.1$.}
\label{fig:distribution}
\end{figure*}

\begin{figure*}[t]
    \centering
    \subfigure[$\hat{q}=0.01$]{
    \begin{minipage}[t]{0.23\textwidth}
    \centering
    \includegraphics[width=1.7in]{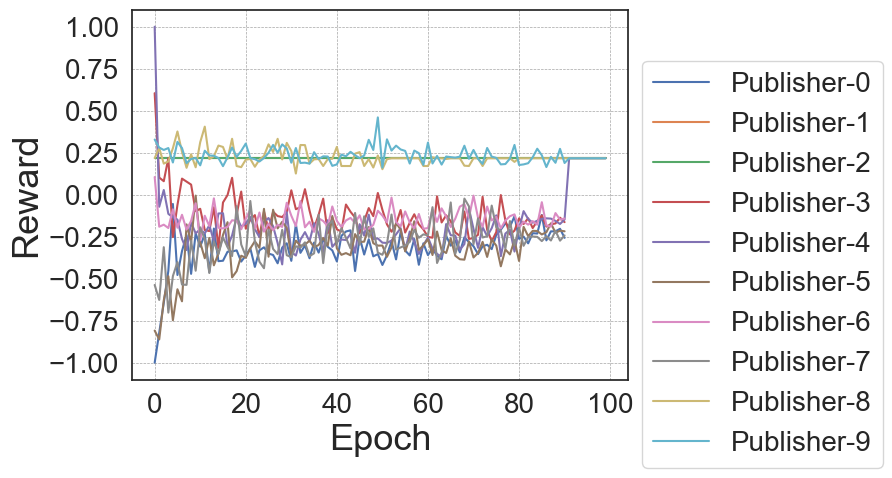}
    \end{minipage}
    \label{fig:initial_raw_interest_1}
    }
    \subfigure[$\hat{q}=0.05$]{
    \begin{minipage}[t]{0.23\textwidth}
    \centering
    \includegraphics[width=1.7in]{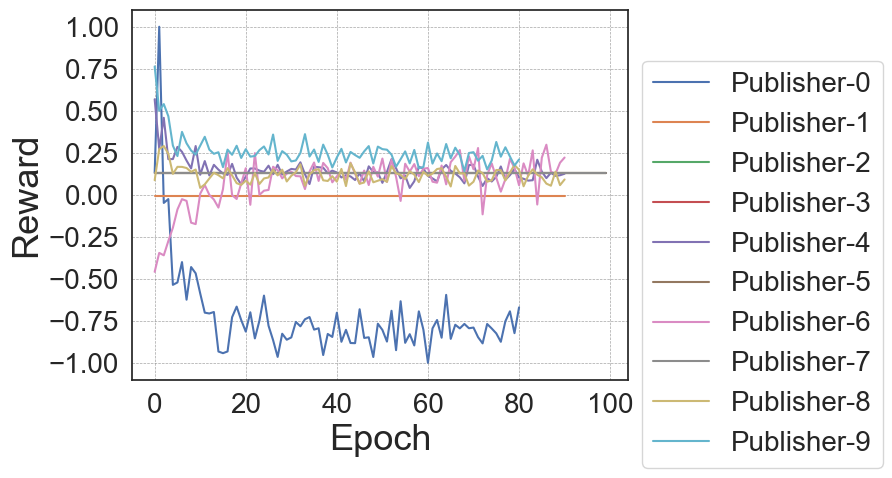}
    \end{minipage}
    \label{fig:initial_raw_interest_2}
    }
    \subfigure[$\hat{q}=0.1$]{
    \begin{minipage}[t]{0.23\textwidth}
    \centering
    \includegraphics[width=1.7in]{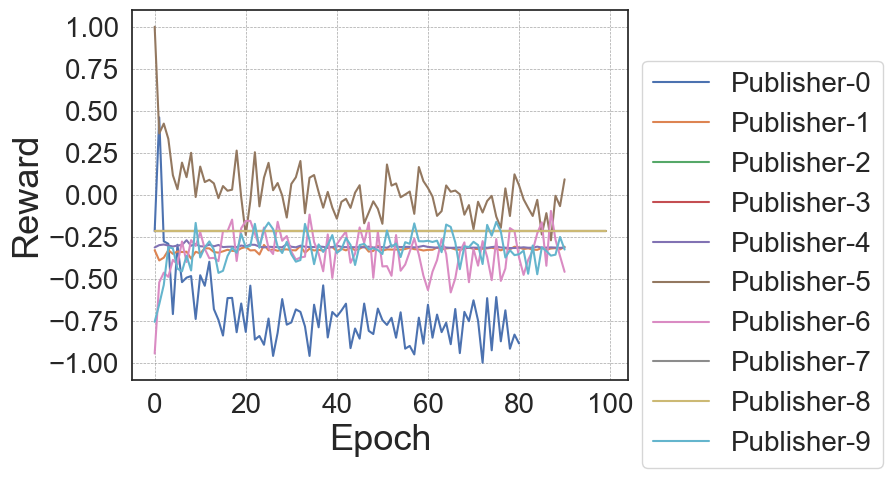}
    \end{minipage}
    \label{fig:initial_raw_interest_3}
    }
    \subfigure[$\hat{q}=0.5$]{
    \begin{minipage}[t]{0.23\textwidth}
    \centering
    \includegraphics[width=1.7in]{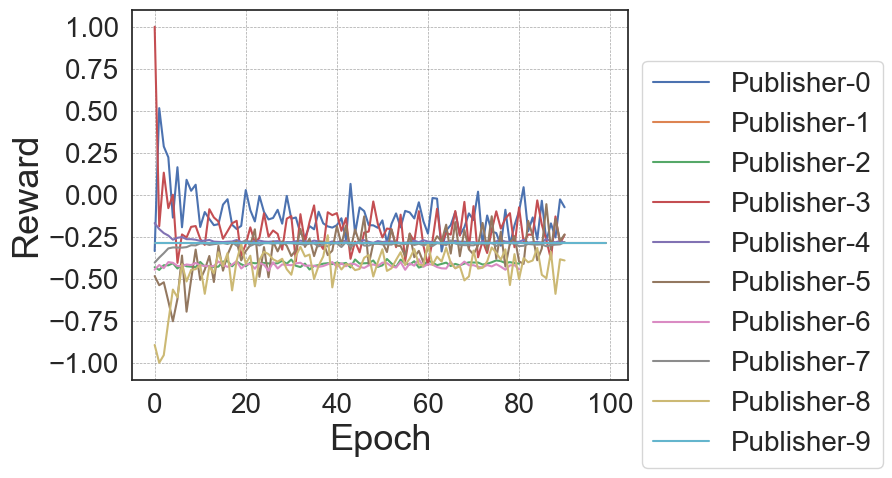}
    \end{minipage}
    \label{fig:initial_raw_interest_4}
    }
    \vspace{-0.2in}
\caption{Comparison between different settings of the \textcolor{teal}{initial raw interest $\hat{q}$} in terms of normalized reward with a setting of $N=10$, uniform-distributed, $|\Omega_{\mathcal{R}_{i,h,\Theta}}|=10$, $\hat{d}=10$, $\hat{\mathcal{I}}=2$, $\hat{\mathcal{O}}=0.1$.}
\label{fig:initial_raw_interest}
\end{figure*}

\begin{figure*}[t]
    \centering
    \subfigure[$|\Omega_{\mathcal{R}_{i,h,\Theta}}|=5$]{
    \begin{minipage}[t]{0.23\textwidth}
    \centering
    \includegraphics[width=1.7in]{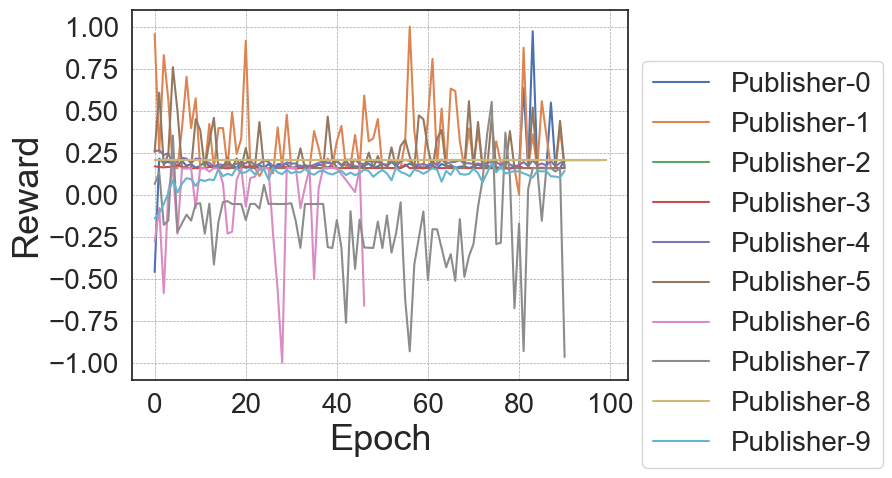}
    \end{minipage}
    \label{fig:candidate_set_1}
    }
    \subfigure[$|\Omega_{\mathcal{R}_{i,h,\Theta}}|=10$]{
    \begin{minipage}[t]{0.23\textwidth}
    \centering
    \includegraphics[width=1.7in]{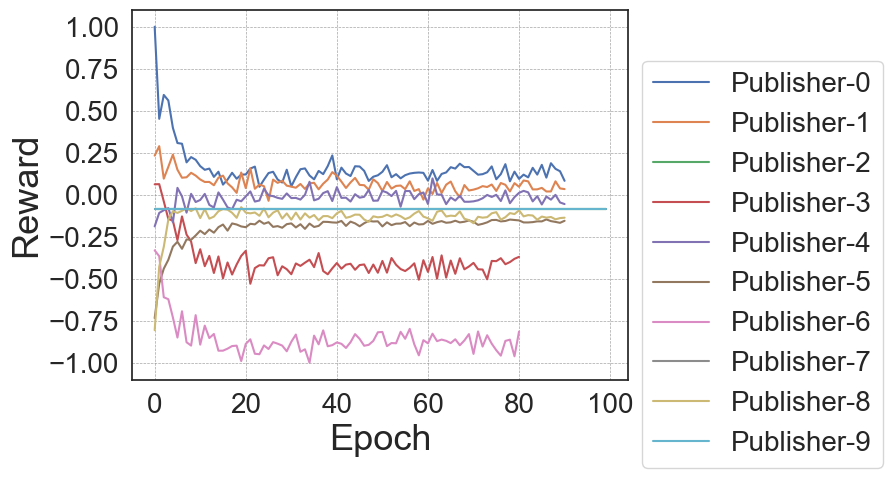}
    \end{minipage}
    \label{fig:candidate_set_2}
    }
    \subfigure[$|\Omega_{\mathcal{R}_{i,h,\Theta}}|=20$]{
    \begin{minipage}[t]{0.23\textwidth}
    \centering
    \includegraphics[width=1.7in]{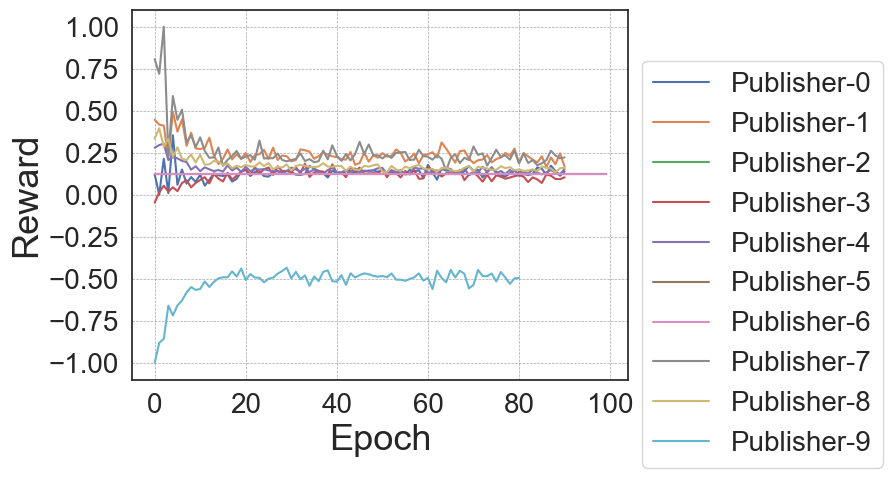}
    \end{minipage}
    \label{fig:candidate_set_3}
    }
    \subfigure[$|\Omega_{\mathcal{R}_{i,h,\Theta}}|=100$]{
    \begin{minipage}[t]{0.23\textwidth}
    \centering
    \includegraphics[width=1.7in]{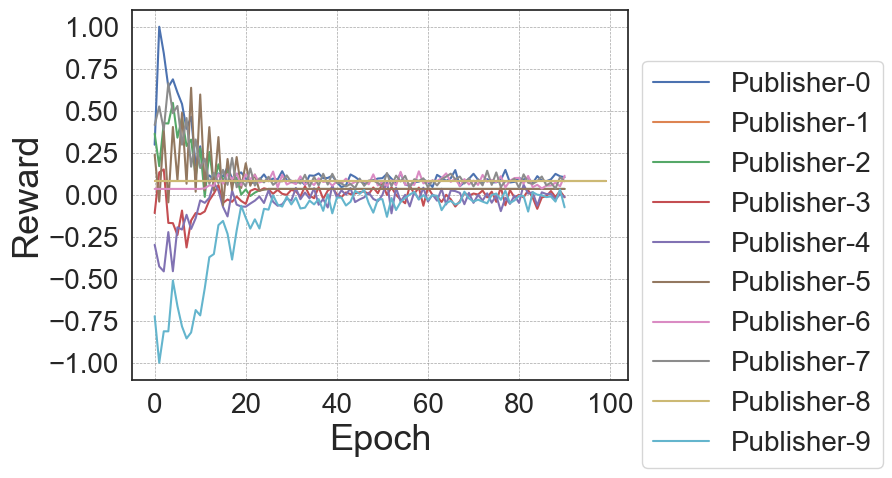}
    \end{minipage}
    \label{fig:candidate_set_4}
    }
    \vspace{-0.2in}
\caption{Comparison between different settings of the \textcolor{teal}{candidate set size $|\Omega_{\mathcal{R}_{i,h,\Theta}}|$} in terms of normalized reward with a setting of $N=10$, uniform-distributed, $\hat{q}=0.01$, $\hat{d}=10$, $\hat{\mathcal{I}}=2$, $\hat{\mathcal{O}}=0.1$.}
\label{fig:candidate_set}
\end{figure*}

\begin{figure*}[t]
    \centering
    \subfigure[$\hat{d}=5$]{
    \begin{minipage}[t]{0.23\textwidth}
    \centering
    \includegraphics[width=1.7in]{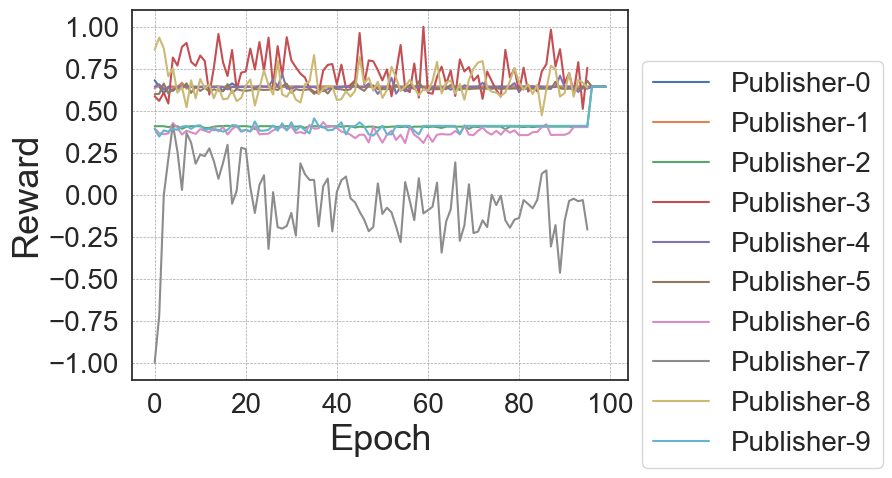}
    \end{minipage}
    \label{fig:initial_decay_1}
    }
    \subfigure[$\hat{d}=10$]{
    \begin{minipage}[t]{0.23\textwidth}
    \centering
    \includegraphics[width=1.7in]{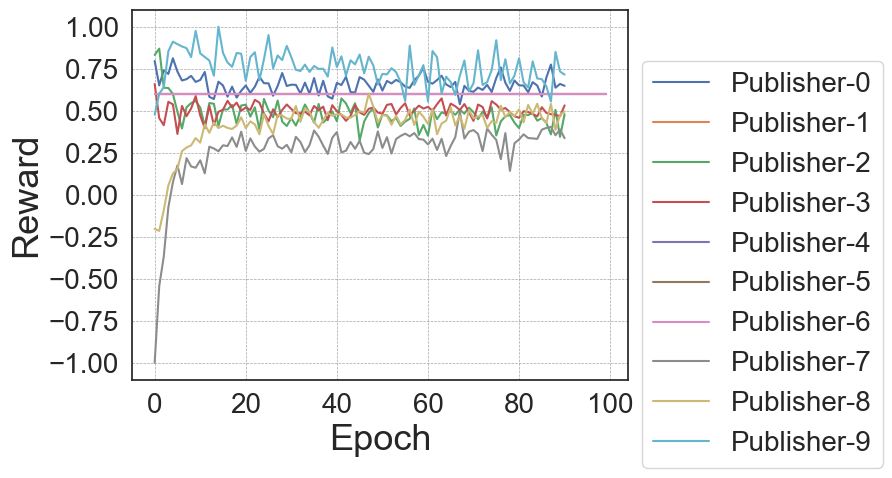}
    \end{minipage}
    \label{fig:initial_decay_2}
    }
    \subfigure[$\hat{d}=20$]{
    \begin{minipage}[t]{0.23\textwidth}
    \centering
    \includegraphics[width=1.7in]{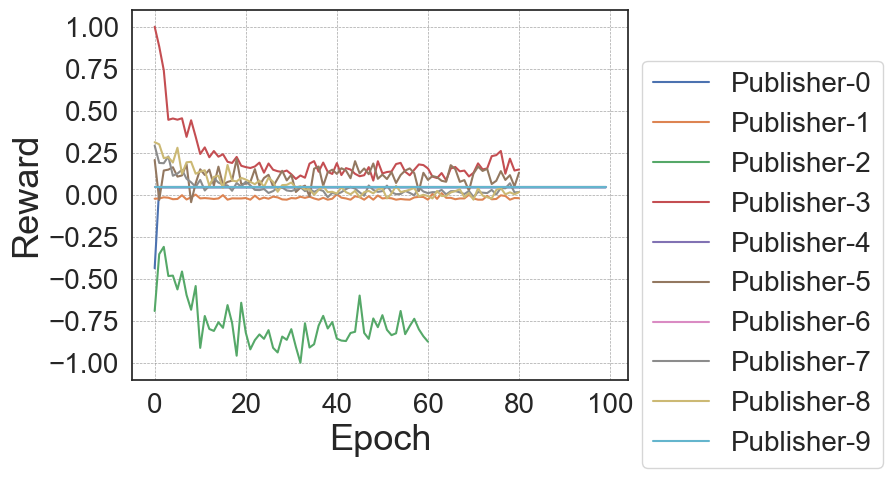}
    \end{minipage}
    \label{fig:initial_decay_3}
    }
    \subfigure[$\hat{d}=30$]{
    \begin{minipage}[t]{0.23\textwidth}
    \centering
    \includegraphics[width=1.7in]{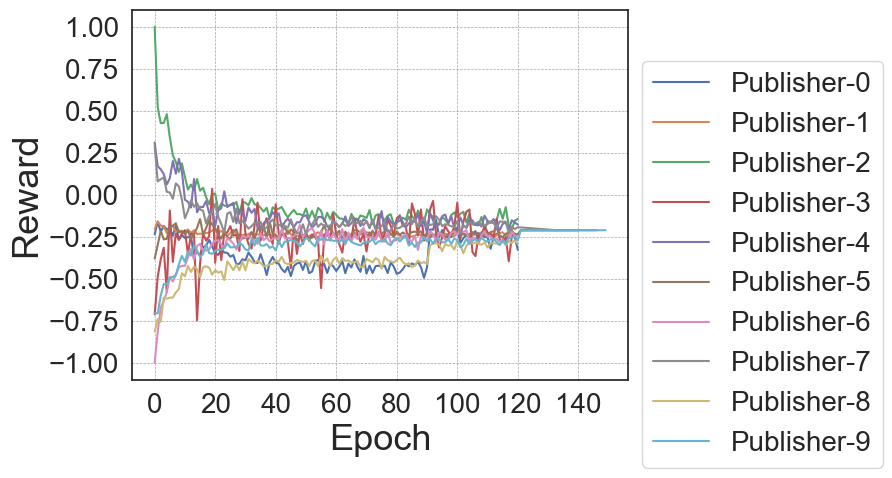}
    \end{minipage}
    \label{fig:initial_decay_4}
    }
        \vspace{-0.2in}
\caption{Comparison between different settings of the \textcolor{teal}{initial decay parameter $\hat{d}$} in terms of normalized reward with a setting of $N=10$, uniform-distributed, $|\Omega_{\mathcal{R}_{i,h,\Theta}}|= 10$, $\hat{q}=0.01$, $\hat{\mathcal{I}}=2$, $\hat{\mathcal{O}}=0.1$.}
\label{fig:initial_decay}
\end{figure*}

\begin{figure*}[t]
    \centering
    \subfigure[$\hat{\mathcal{I}}=2$]{
    \begin{minipage}[t]{0.23\textwidth}
    \centering
    \includegraphics[width=1.7in]{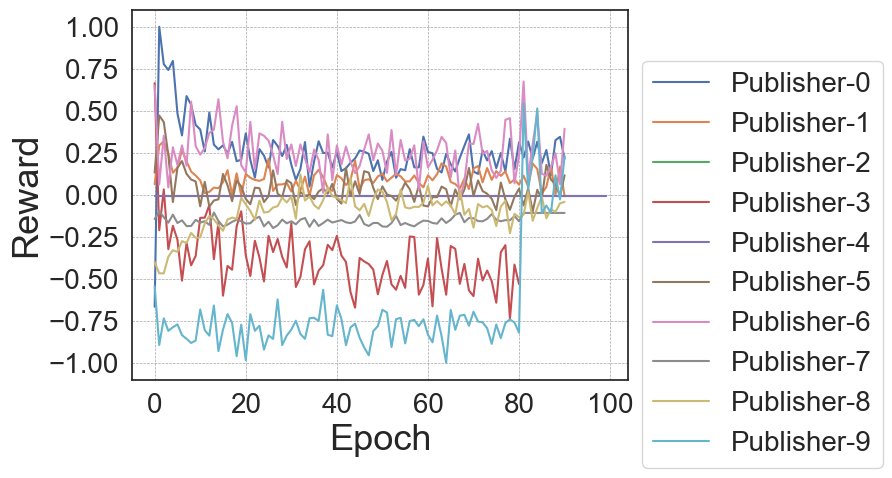}
    \end{minipage}
    \label{fig:initial_fixed_reward_1}
    }
    \subfigure[$\hat{\mathcal{I}}=4$]{
    \begin{minipage}[t]{0.23\textwidth}
    \centering
    \includegraphics[width=1.7in]{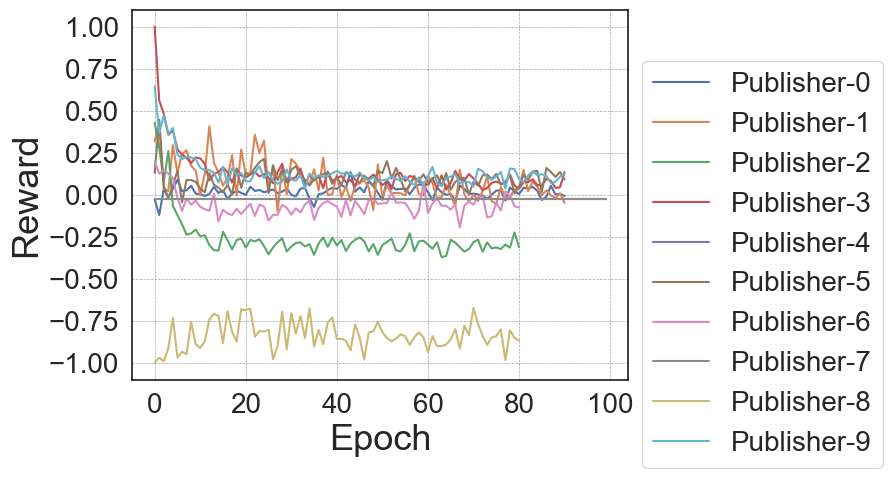}
    \end{minipage}
    \label{fig:initial_fixed_reward_2}
    }
    \subfigure[$\hat{\mathcal{I}}=8$]{
    \begin{minipage}[t]{0.23\textwidth}
    \centering
    \includegraphics[width=1.7in]{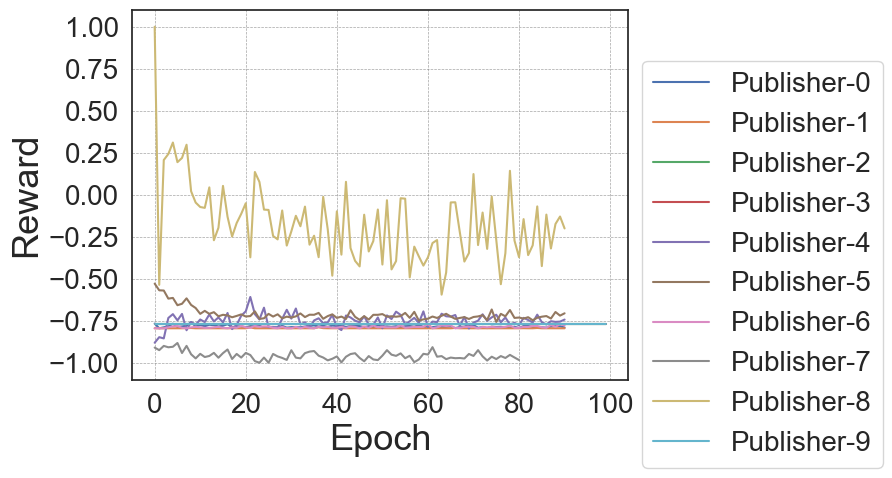}
    \end{minipage}
    \label{fig:initial_fixed_reward_3}
    }
    \subfigure[$\hat{\mathcal{I}}=16$]{
    \begin{minipage}[t]{0.23\textwidth}
    \centering
    \includegraphics[width=1.7in]{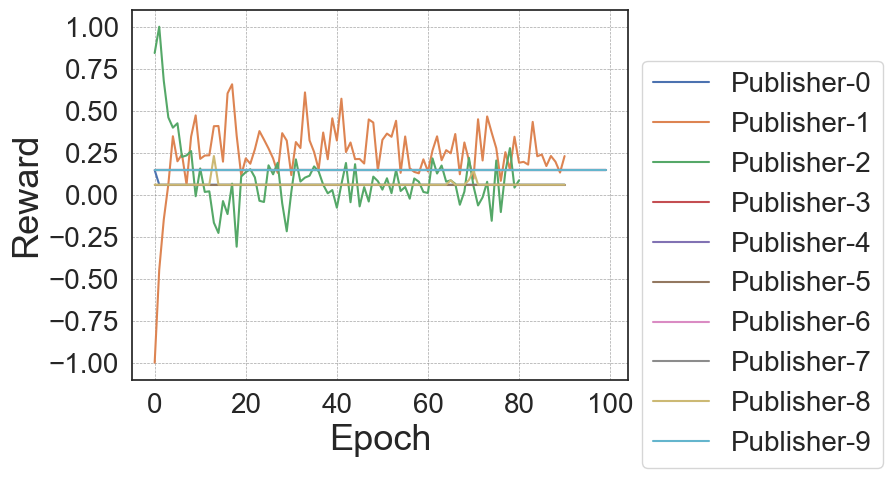}
    \end{minipage}
    \label{fig:initial_fixed_reward_4}
    }
    \vspace{-0.2in}
\caption{Comparison between different settings of the \textcolor{teal}{initial fixed reward $\hat{\mathcal{I}}$} in terms of normalized reward with a setting of $N=10$, uniform-distributed, $|\Omega_{\mathcal{R}_{i,h,\Theta}}|= 10$, $\hat{q}=0.01$, $\hat{d}=10$, $\hat{\mathcal{O}}=0.1$.}
\label{fig:initial_fixed_reward}
\end{figure*}

\begin{figure*}[t]
    \centering
    \subfigure[$\hat{\mathcal{O}}=0.1$]{
    \begin{minipage}[t]{0.23\textwidth}
    \centering
    \includegraphics[width=1.7in]{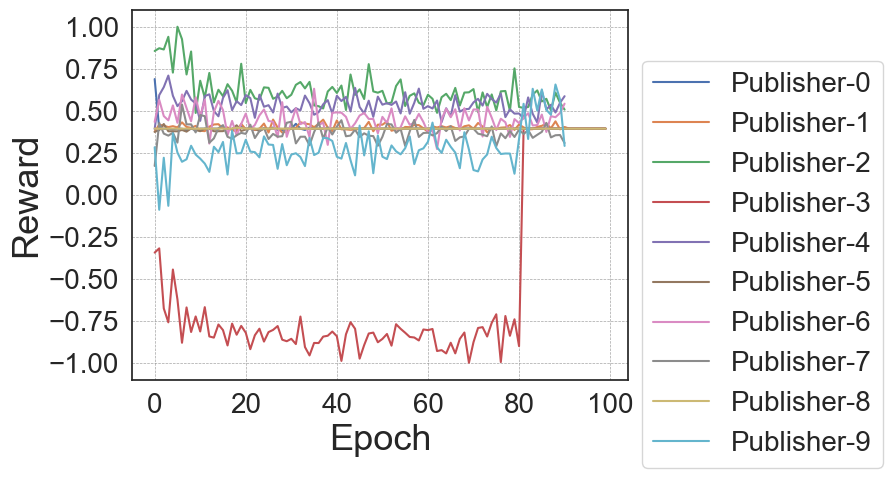}
    \end{minipage}
    \label{fig:initial_fixed_expense_1}
    }
    \subfigure[$\hat{\mathcal{O}}=0.2$]{
    \begin{minipage}[t]{0.23\textwidth}
    \centering
    \includegraphics[width=1.7in]{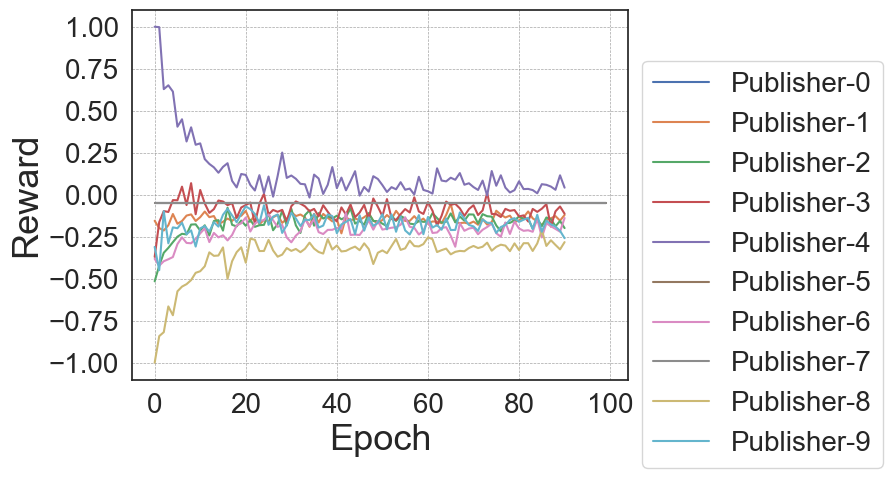}
    \end{minipage}
    \label{fig:initial_fixed_expense_2}
    }
    \subfigure[$\hat{\mathcal{O}}=0.4$]{
    \begin{minipage}[t]{0.23\textwidth}
    \centering
    \includegraphics[width=1.7in]{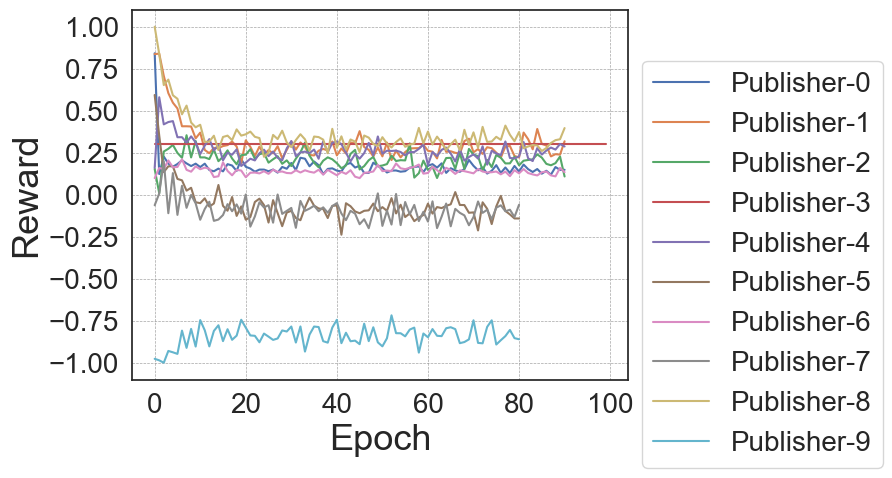}
    \end{minipage}
    \label{fig:initial_fixed_expense_3}
    }
    \subfigure[$\hat{\mathcal{O}}=0.8$]{
    \begin{minipage}[t]{0.23\textwidth}
    \centering
    \includegraphics[width=1.7in]{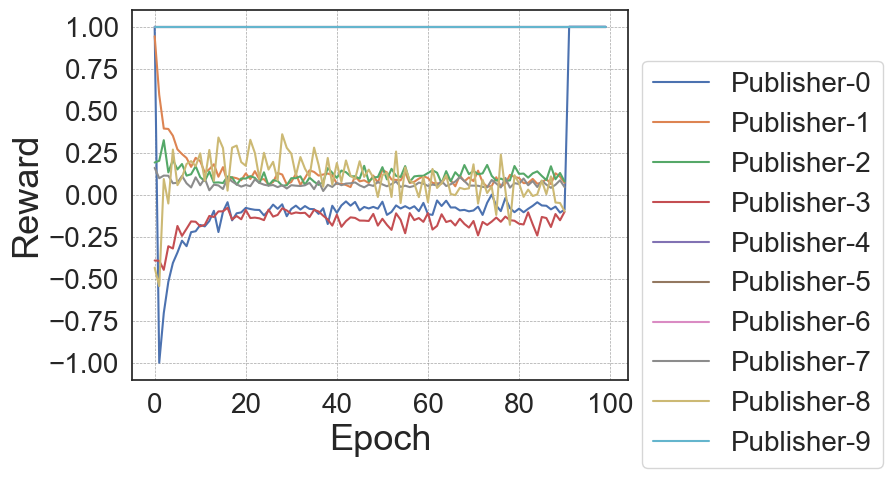}
    \end{minipage}
    \label{fig:initial_fixed_expense_4}
    }
     \vspace{-0.2in}
\caption{Comparison between different settings of the \textcolor{teal}{initial fixed expense $\hat{\mathcal{O}}$} in terms of normalized reward with a setting of $N=10$, uniform-distributed, $|\Omega_{\mathcal{R}_{i,h,\Theta}}|= 10$, $\hat{q}=0.01$, $\hat{d}=10$, $\hat{\mathcal{I}}=2$.}
\label{fig:initial_fixed_expense}
\end{figure*}

The rewards in all figures (Fig.\ref{fig:publisher_num}-Fig.\ref{fig:initial_fixed_expense}) are normalized to keep the reward values in a specific convenient range and obtain the best value approximation since the value distribution function could significantly affect the gradients~\cite{c51-settings}. The occurrence of fewer result values for a particular setting across all epochs can be attributed to the dynamics of the ML resource marketplace, where the income from published NFTs is only realized at the end of their expiry height $d$. If an NFT fails to be referenced before the maximum epoch, its reward value is set to null.

Fig.\ref{fig:publisher_num} illustrates the convergence trend of rewards across $100$ epochs for various configurations of publisher numbers. The analysis reveals that as the number of publishers increases from $10$ to $40$, their rewards tend to converge. The majority of publishers settle at a moderate reward level, approximately around the normalized reward of $0$. Only a small fraction achieve high normalized rewards close to $1$, and similarly, few reach low normalized rewards of $-1$. This trend becomes more pronounced with an increase in the number of publishers, suggesting that employing a DRL-based approach in the pricing game results in relatively fair outcomes when the number of publishers is reasonable and resources are approximately uniformly distributed.

Fig.\ref{fig:distribution} shows how different statistical distributions impact the rewards of initial NFTs. In the uniform distribution, we observe a general equilibrium among publishers, with most displaying similar rewards and a few outliers adjusting to fit the common trend, which can be found in Figs.\ref{fig:publisher_num_10},\ref{fig:uniform_distribution},\ref{fig:initial_raw_interest_1},\ref{fig:candidate_set_2},\ref{fig:initial_decay_2},\ref{fig:initial_fixed_reward_1},\ref{fig:initial_fixed_expense_1} under the same setting. 
The normal distribution aligns with the expected bell-curve shape, showing a central clustering of rewards.
In contrast, the Pareto distribution deviates from the anticipated 80-20 rule, with most publishers having uniform-like rewards and only a few falling behind, indicating a lesser effect of resource concentration. The Poisson distribution, however, presents a more varied pattern with noticeable fluctuations among publishers, while collectively maintaining a reward range between -0.5 and +0.5. This unexpectedly close alignment to the uniform distribution, but with even more balance, suggests our system's proficiency in managing the randomness of the Poisson distribution and maintaining equilibrium.

Fig.\ref{fig:initial_raw_interest} illustrates the impact of different initial raw interest rates ($\hat{q}$) on publisher behaviors. At $\hat{q}=0.01$, we observe moderate fluctuations, with most publishers settling around a -0.25 reward. Increasing the rate to 0.05 results in a stabilization of rewards around 0.2, though some publishers choose not to participate, likely to avoid costs. At a higher rate of 0.1, similar behaviors are evident, but with an approximate decrease in rewards from 0.2 to -0.2, indicating cost-conscious strategies. At the highest rate of 0.5, there is a marked convergence of rewards within -0.5 to -0.2, showing a collectively cautious approach in response to higher costs. It suggests that as financial risks increase, publishers tend to focus more on minimizing losses than maximizing gains, demonstrating the significant role of economic factors in shaping strategies within NFT ecosystems.

Fig.\ref{fig:candidate_set} and Fig.\ref{fig:initial_decay} collectively demonstrate the influence of candidate set size $|\Omega_{\mathcal{R}_{i,h,\Theta}}|$ and initial decay parameter $\hat{d}$ on publishers' reward patterns. In both scenarios, with smaller candidate set sizes or lower decay parameters, there's a significant fluctuation in rewards among some publishers, indicative of a highly competitive or unbalanced environment. This volatility is particularly pronounced when the candidate set size is limited to $|\Omega_{\mathcal{R}_{i,h,\Theta}}|$ or $\hat{d}$ is low, reflecting the intense competition and rapid shifts in the landscape of reference fees and incentives due to shorter effective NFT durations. As both the candidate set size and $\hat{d}$ increase, these fluctuations decrease, leading to stabilization and convergence of rewards. This pattern suggests a movement towards a more equitable distribution, mirroring the expected outcomes of a uniform distribution. The increase in diversity and options within the candidate set, along with the extended lifespan of NFTs due to a higher decay parameter, both contribute to a fairer, more predictable, and harmonious environment for publishers, emphasizing the crucial role these parameters play in shaping the dynamics of competitive interactions and reward distribution in NFT ecosystems.

Fig.\ref{fig:initial_fixed_reward} and Fig.\ref{fig:initial_fixed_expense} reveal the interplay between initial fixed reward $\hat{\mathcal{I}}$ and initial fixed expense $\hat{\mathcal{O}}$, and their impact on publishers' reward behaviors in NFT ecosystems. The initial fixed reward $\hat{\mathcal{I}}$, granted for superior performance in newly published NFTs, induces notable variations in rewards among publishers when set low. However, as $\hat{\mathcal{I}}$ increases, these fluctuations diminish, leading to a more uniform reward distribution. Interestingly, at higher $\hat{\mathcal{I}}$ values, most publishers opt for inactivity, reflected in consistently zero rewards, with only a few active publishers earning similar rewards, indicating a strategic shift towards non-participation in publishing due to the high reward benchmark. 
Conversely, the initial fixed expense $\hat{\mathcal{O}}$, a preventive measure against malicious attacks, shows a less pronounced impact on publisher behavior at lower levels. Yet, as $\hat{\mathcal{O}}$ escalates, it becomes a critical factor, potentially fostering monopolistic trends within the NFT space, as evidenced in Fig.~\ref{fig:initial_fixed_expense_4}. This suggests a delicate balance in setting $\hat{\mathcal{O}}$: too low, and it is ineffectual as a deterrent; too high, and it risks centralizing power and stifling diversity in publishing. These observations underline the need for judicious calibration of $\hat{\mathcal{I}}$ and $\hat{\mathcal{O}}$ to maintain a dynamic, competitive, yet fair NFT environment, encouraging active participation while safeguarding against disruptive behaviors.

\smallskip
\noindent\textbf{Takeaways.} We provide the following key experimental results.

\textit{\ding{172} Adaptability in diverse distribution environments.} Our system demonstrates remarkable adaptability and resilience across different initial NFT quality distributions, effectively managing variations in resource concentration. The deviation from the 80-20 rule in Pareto distributions and the unexpected equilibrium in Poisson distributions underline the system's capacity to maintain balance and fairness, irrespective of the underlying distribution characteristics.

\textit{\ding{173} The Interplay of economic factors and publisher behavior.} Economic factors (e.g., initial raw interest rate, fixed expenses) significantly influence publisher strategies within NFT ecosystems. As financial risks heighten, publishers focus more on risk mitigation than gain maximization, leading to strategic shifts like reduced participation or monopoly formation.

\textit{\ding{174} Equilibrium and fairness in reward distribution.} Increasing the number of publishers, along with adjustments in candidate set size and NFT lifespan, contributes to a more equitable reward distribution. This phenomenon aligns with uniform distribution principles, suggesting that a larger, more diverse candidate set and extended NFT lifespans promote a fairer and more stable environment.

\section{Discussion}

\subsection{Additional Capabilities}\label{subsec-capability}

\noindent\textbf{Flexibility for customized requirements.}
The considered ML resource marketplace accommodates the provision of additional customized services. Users registered with the marketplace might be interested in customized preprocessing or privacy-sensitive datasets and models. Such preferences can be related to various attributes, such as the quality of training resources, fairness, bias mitigation, privacy enforcement, and accountability. Users can optionally request all these aspects for an additional charge. These attributes can also enhance the coherence of regulatory compliance.

Mathematically, we represent the normalized service demands as \(\varphi_{\dagger}\in[0,1]\), and the unit price of offering customized services as \(\psi_{\dagger} \in [0, \bar{\psi_{\dagger}}]\). This leads to two updates to the system:
$\mathcal{A}_{\text{out}} \Longleftarrow\varphi_{\dagger} \cup \mathcal{A}_{\text{out}}$ and
$\mathcal{A}_{\text{in}}  \Longleftarrow\psi_{\dagger} \cup \mathcal{A}_{\text{in}}$.


\noindent\textbf{Adaptability for multiple conditions.}
The reference incentive model seamlessly interplays with the marketplace's dynamics. The model's resilience in various NFT quality distributions showcases its ability to adapt to different market conditions, maintaining fairness and resource balance. This adaptability is critical in managing the economic factors influencing publisher strategies within the NFT ecosystem. As financial risks escalate, the model's framework can further accommodate strategic shifts in publisher behavior, emphasizing the importance of setting balanced economic parameters.

\smallskip
\noindent\textbf{Fairness in resource/profit distribution.} The model's approach to reward distribution, by increasing publisher numbers and adjusting NFT parameters, aligns with uniform distribution principles. This strategy ensures equitable rewards and fosters a stable environment, highlighting the model's role in sustaining a harmonious and competitive NFT marketplace.

\subsection{Broader Adoption}\label{subsec-application}

\noindent\textbf{Compatibility with NFT standards.} The proposed framework demonstrates extensive compatibility with various NFT standards, enhancing its adoption potential in diverse scenarios:
\begin{itemize}
    \item \textit{Composability and hierarchical relationships:} EIP-\hlhref{https://eips.ethereum.org/EIPS/eip-998}{998}, \hlhref{https://eips.ethereum.org/EIPS/eip-6150}{6150}, and \hlhref{https://eips.ethereum.org/EIPS/eip-7510}{7510} focus on composability and hierarchical NFT structure. Our model complements these by effectively managing and tracking the complex inter-NFT relationships, which enriches the depth of interactions within these frameworks.

    \item \textit{Utility and ownership:} EIP-\hlhref{https://eips.ethereum.org/EIPS/eip-5006}{5006}, \hlhref{https://eips.ethereum.org/EIPS/eip-6059}{6059}, and \hlhref{https://eips.ethereum.org/EIPS/eip-5528}{5528} emphasize flexible utility, ownership models, and financial transactions in NFTs. Our model can augment these standards by providing robust mechanisms for tracking ownership and utility changes and facilitating transparent financial dealings.

    \item \textit{Extended functionalities:} EIP-\hlhref{https://eips.ethereum.org/EIPS/eip-4519}{4519}, \hlhref{https://eips.ethereum.org/EIPS/eip-5023}{5023}, and \hlhref{https://eips.ethereum.org/EIPS/eip-5606}{5606} introduce use cases such as physical asset linkage and multiverse asset management. Our model can enhance these use cases by offering a structured system for tracking and incentivizing interactions, thus adding value to these applications.

    \item \textit{External standards:} While our design originates from the Ethereum ecosystem, our overarching reference incentive model can be extended to L2 NFTs on other public chains that incorporate reference functions in the future.

\end{itemize}

\smallskip
\noindent\textbf{Integration with NFT Marketplaces.} Our model allows straightforward integration into platforms such as OpenSea, Rarible, and Binance, enriching their diverse collections with advanced relationship tracking and analysis capabilities. This integration is designed to be unobtrusive, ensuring the maintenance of existing marketplace functionalities and user experiences. Users have a strong inherent motivation to earn continuous profits. This model ensures a transparent and efficient way to track and reward contributions within the NFT ecosystem and amplifies their profit potential from sharing NFTs and data. Our approach aligns well with the evolving needs of the digital asset community, offering a balanced advantage for both marketplace operators and users. 

\smallskip
\noindent\textbf{Extension to any resource marketplace.} As discussed in the case study, our model can be applied to any (ML-)resource marketplace scenario. Here, models and datasets are openly shared and developed through mutual referencing, forming a DAG-based network within the marketplace. However, the framework's potential extends beyond the resource marketplace to other areas with different complexities. For instance, it can be used for joint ventures involving data-driven investment strategies in finance, which creates a transparent network of contributions and rewards. Similarly, in the metaverse, this model can be crucial for managing the creation and sharing of virtual assets. It can trace and incentivize contributions in the virtual economy, which ensures fair reward distribution among digital artists, developers, and users.


\section{Related Work}\label{sec-rw}

\noindent\textbf{NFT-related standards.} Since the advent of NFTs in 2021~\cite{wang2021non}, a plethora of sub-domains have emerged, consisting of new applications~\cite{vidal2022new}\cite{wang2022exploring}\cite{cetinic2022understanding}, system security~\cite{das2022understanding}, and addressing anti-law activities such as wash trading~\cite{wen2023nftdisk} and rugpulls~\cite{huang2023miracle}. Instead of relying on nebulous topic classifications, we orthogonally explore the foundational elements — the token standards. Specifically, we review all up-to-date (Jan. 2024) NFT-related token standards within the Ethereum ecosystem (cf. Table~\ref{tab-tokenstandard}), selectively presenting their primary functions, distinctive features, and potential applications. To our surprise, the landscape of NFT standards has experienced a remarkable expansion, skyrocketing from approx. 10 (mid-2021) to 50+, surpassing the pace observed in other standard tracks that are assumed to progress at a slower rate.

\smallskip
\noindent\textbf{Recent incentive analysis.} The recent wave of incentive analyses has predominantly focused on the Ethereum ecosystem and PoS protocols following its transition.  Deb et al.~\cite{possafety2023} proposed a model for analyzing the cryptoeconomic safety of PoS blockchain (Ethereum 2.0) and observed the bounds between cost and profit from malicious corruption.  Cortes et al.~\cite{cortes2023autopsy} systematically studied Ethereum's post-merge incentive mechanism and empirically analyzed the performance of the corresponding network. Huang et al.~\cite{huang2021rich} observed the diverse incentive models implemented in current PoS blockchains and provided an in-depth analysis of various PoS representatives.
More incentive studies also apply to broader areas such as DeFi products (auctions~\cite{wu2023strategic}\cite{bahrani2023bidders}, AMM~\cite{kulkarni2022towards}\cite{bartoletti2022theory}, DEX~\cite{hasbrouck2023economic}), MEV~\cite{chionas2023gets}, transaction fees (post-EIP1559, \cite{ferreira2021dynamic}\cite{leonardos2023optimality}), NFTs~\cite{heimbach2023defi}\cite{chen2022absnft}\cite{bao2023herding}, etc.

\smallskip
\noindent\textbf{DAG formation in L1.} DAG topology has been widely applied in blockchain systems~\cite{wang2023sok} for improving scalability and performance \cite{wang2021weak}\cite{yu2020ohie}\cite{bagaria2019prism}\cite{benvcic2018distributed}, followed by a cascade of supplementary analyses, including security~\cite{amores2023we}\cite{kiayias2019trees}\cite{wang2020security}, fairness~\cite{raikwar2023fairness}, simulation \cite{zander2019dagsim}\cite{lin2023tanglesim} and optimizations/protection \cite{malkhi2022maximal}. Most L1 protocols utilize DAG to reshape their data structures \cite{silvano2020iota}\cite{churyumov2016byteball}\cite{baird2016swirlds}\cite{li2020decentralized} and reform consensus progress \cite{keidar2023cordial}\cite{keidar2021all}\cite{schett2021embedding}\cite{danezis2022narwhal}\cite{spiegelman2022bullshark}, which requires highly adaptive modifications to their corresponding foundational systems.

\smallskip
\noindent\textbf{DAG adoption for L2.} Diverging from applying DAG to blockchain structure/consensus, several studies have embraced the DAG philosophy to construct upper-layer decentralized applications that operate independently of their L1 counterparts. Wang et al.~\cite{wang2023referable} introduced the utilization of DAG topology in building an NFT-rewarding market, as expounded in this paper. Yu et al.~\cite{10174882} stepped further by leveraging Graph Neural Networks (GNNs) to develop a prediction and recommendation framework. Simultaneously, DAGs can also be used in the closely related domain of decentralized federated learning, aiding in the parallel processing of task workers~\cite{ironforge}.

\smallskip
\noindent\textbf{Game theory in blockchain.} Various game theory models have found application in blockchain systems~\cite{liu2019survey} such as utilizing stochastic games~\cite{kiayias2016blockchain}, cooperative games~\cite{lewenberg2015bitcoin}, evolutionary games~\cite{kim2019mining}, Stackelberg games~\cite{chen2022absnft}, subgame perfect Nash equilibrium~\cite{qin2022bdts}. They serve to analyze practical scenarios at different levels, such as  staking/mining pools \cite{gersbach2022staking}\cite{zhang2022insightful}\cite{li2020mining}\cite{wang2019pool}, mining participation~\cite{chaidos2023blockchain},  validator selection \cite{zhang2023rationally}\cite{gavzi2023fait}, mining behaviors \cite{eyal2015miner}\cite{bai2021blockchain}\cite{cheung2021griefing}, off-chain transactions \cite{rain2023towards} and adversarial strategies \cite{estradabreaking}\cite{heimbach2022eliminating}\cite{negy2020selfish}\cite{kwon2017selfish}.

\smallskip
\noindent\textbf{DRL-assisted applications.} Many areas realize the superiority of applying deep reinforcement learning to solve complex problems, including blockchain L1 solutions~\cite{10201805, zhang2024tbdd}, IoT and telecommunication~\cite{8657779,8231220}, microchip design~\cite{liang2020adaptive}, robotic design~
\cite{gu2017deep}, gaming~\cite{lample2017playing}, healthcare~\cite{yu2021reinforcement}, economics and finance~\cite{Charpentier2023, math8101640}, training/jailbreaking ML models~\cite{hu2023enabling, zou2023universal} and more~\cite{wang2022deep}.
\section{Conclusion}\label{sec-conclusion}

In this paper, we study how to make NFT rewards better. After checking existing NFT standards and projects, we noticed that current reward systems are one-time and separate. In response, we propose a \textit{reference incentive} model, where NFTs are structured as DAG, for deep analyses. Our model allows each token to grow its connections, earning more rewards from future transactions. We also prove that this model works better both in theory and practice.


\normalem
\bibliographystyle{unsrt}
\bibliography{bib.bib}


\appendix

\begin{table*}[!hbtp]
 \caption{Ethereum NFT-related Standards: Categorized by \textit{Final}-\textit{LastCall}-\textit{Review}-\textit{Draft} (Updated on Feb. 2023)} 
 \label{tab-tokenstandard}
  \centering
  \renewcommand\arraystretch{1}
 \resizebox{\linewidth}{!}{
 \begin{tabular}{clllr}
    \toprule
    \multicolumn{1}{c}{\textbf{EIP-}}  
     & \multicolumn{1}{c}{\textbf{Title}} &  \multicolumn{1}{c}{\textbf{Main (\textcolor{teal}{new}) functions/events/metadata}} & \multicolumn{1}{c}{\textbf{Feature}} &
    \multicolumn{1}{c}{\textbf{Application}}  \\
    \midrule
    \hlhref{https://eips.ethereum.org/EIPS/eip-721}{721} & Non-fungible token & \textcolor{teal}{$\mathsf{tokenID}$} & Artwork/IP    \\
    \hlhref{https://eips.ethereum.org/EIPS/eip-1155}{1155} &  Multi token standards & \textcolor{teal}{$\mathsf{batchProcessing}$} & Adding attributes for groups &   Game  \\
    
    \hlhref{https://eips.ethereum.org/EIPS/eip-2309}{2309} & ERC-721 Consecutive Transfer Extension & \textcolor{teal}{$\mathsf{ConsecutiveTransfer}$}& Upgrading events & Authorization \\
    \hlhref{https://eips.ethereum.org/EIPS/eip-2981}{2981} & NFT Royalty Standard & \textcolor{teal}{$\mathsf{royaltyInfo}$} & Retrieving the royalty payment infomation & Royalty payments \\
    \hlhref{https://eips.ethereum.org/EIPS/eip-3525}{3525}   & Semi-Fungible Token & \textcolor{teal}{$\mathsf{SlotChanged}$}, \textcolor{teal}{$\mathsf{slotOf}$}& Additional attribute for semi-fungible & Financial market  \\
   \hlhref{https://eips.ethereum.org/EIPS/eip-4519}{4519}    & Non-Fungible Tokens Tied to Physical Assets & \textcolor{teal}{$\mathsf{ownerEngagement}$}, \textcolor{teal}{$\mathsf{userEngagement}$} & Representing physical assets& IoT indusrty\\
    \hlhref{https://eips.ethereum.org/EIPS/eip-4906}{4906}    & EIP-721 Metadata Update Extension &  \textcolor{teal}{$\mathsf{MetadataUpdate}$} & Upgrading events \\
    \hlhref{https://eips.ethereum.org/EIPS/eip-4907}{4907} & 	Rental NFT, an Extension of EIP-721   & \textcolor{teal}{$\mathsf{userExpires}$}& Adding a new role and timer &   Rental market \\
    \hlhref{https://eips.ethereum.org/EIPS/eip-4910}{4910} & Royalty Bearing NFTs & CRUD (\textcolor{teal}{RoyaltyAccount}), \textcolor{teal}{$\mathsf{royaltyPayOut}$} & Upgrading royalty account management & Royalty payment \\
    \hlhref{https://eips.ethereum.org/EIPS/eip-4955}{4955} & Vendor Metadata Extension for NFTs &\textcolor{teal}{namespaces} & Upgrading metadata & Metaverse 3D modelling \\
    \hlhref{https://eips.ethereum.org/EIPS/eip-5006}{5006}  & Rental NFT, NFT User Extension & \textcolor{teal}{UserRecord} & Adding the new role of user & Rental Market   \\
    \hlhref{https://eips.ethereum.org/EIPS/eip-5007}{5007} & Time NFT, ERC-721 Time Extension & \textcolor{teal}{startTime}, \textcolor{teal}{endTime}& On-chain time management & Lending market  \\
    \hlhref{https://eips.ethereum.org/EIPS/eip-5023}{5023} & Shareable Non-Fungible Token & \textcolor{teal}{share}& Enabling assets to be sharable &  Collaborative projects\\
    \hlhref{https://eips.ethereum.org/EIPS/eip-5192}{5192} & Minimal Soulbound NFTs & \textcolor{teal}{locked}& Bound to a single account  & Soulbound Items  \\
    \hlhref{https://eips.ethereum.org/EIPS/eip-5375}{5375} &  	NFT Author Information and Consent & \textcolor{teal}{authorInfo}& Adding authorship and consent& Authorization  \\
    \hlhref{https://eips.ethereum.org/EIPS/eip-5380}{5380} & 	ERC-721 Entitlement Extension  &\textcolor{teal}{entitle}, \textcolor{teal}{entitlementOf} & Adding entitlement for users & Rental market \\
     \hlhref{https://eips.ethereum.org/EIPS/eip-5489}{5489} &	NFT Hyperlink Extension & \textcolor{teal}{authorizeSlotTo}, \textcolor{teal}{revokeAuthorization}& Adding hyperlinks &  \\
     \hlhref{https://eips.ethereum.org/EIPS/eip-5528}{5528} &	Refundable Fungible Token &escrow (\textcolor{teal}{Fund}/\textcolor{teal}{Refund}/\textcolor{teal}{Withdraw}) & Enabling refund & Rental market \\
     \hlhref{https://eips.ethereum.org/EIPS/eip-5570}{5570} &	Digital Receipt Non-Fungible Tokens & &Adding digital receipts for physical purchases &  \\
     \hlhref{https://eips.ethereum.org/EIPS/eip-5606}{5606} &	Multiverse NFTs &\textcolor{teal}{bundle}, \textcolor{teal}{delegateTokens} & Enabling one-to-many asset mapping & Metaverse, Game \\
     \hlhref{https://eips.ethereum.org/EIPS/eip-5615}{5615} &	ERC-1155 Supply Extension & \textcolor{teal}{totalSupply} & Fetching token supply data &  \\
     \hlhref{https://eips.ethereum.org/EIPS/eip-5725}{5725} &	Transferable Vesting NFT & \textcolor{teal}{claimedPayout}, \textcolor{teal}{vestedPayout} & Vesting tokens & Financial market \\
     \hlhref{https://eips.ethereum.org/EIPS/eip-5773}{5773} &	Context-Dependent Multi-Asset Tokens &\textcolor{teal}{acceptAsset}, \textcolor{teal}{acceptAsset}, \textcolor{teal}{setPriority} & Adding context-dependent output&   \\
     \hlhref{https://eips.ethereum.org/EIPS/eip-6059}{6059} &	Parent-Governed Nestable Non-Fungible Tokens & \textcolor{teal}{DirectOwner}, \textcolor{teal}{Child} & Adding parent-governed nestable relationship & File system   \\
     \hlhref{https://eips.ethereum.org/EIPS/eip-6066}{6066} &	Signature Validation Method for NFTs & \textcolor{teal}{sign}, \textcolor{teal}{isValidSignature} & Verifying signature & E-voting  \\
     \hlhref{https://eips.ethereum.org/EIPS/eip-6105}{6105} &	No Intermediary NFT Trading Protocol &\textcolor{teal}{listItem}, \textcolor{teal}{buyItem} & Adding a marketplace functionality & Marketplace  \\
     \hlhref{https://eips.ethereum.org/EIPS/eip-6147}{6147} &	Guard of NFT/SBT, an Extension of ERC-721 & \textcolor{teal}{changeGuard}, \textcolor{teal}{transferAndRemove} & Adding a new role for new management scheme & Soulbound Items   \\
     \hlhref{https://eips.ethereum.org/EIPS/eip-6150}{6150} &	Hierarchical NFTs & \textcolor{teal}{parentOf}, \textcolor{teal}{childrenOf} & Adding hierarchical structure & File system   \\
     \hlhref{https://eips.ethereum.org/EIPS/eip-6220}{6220} &	Composable NFTs utilizing Equippable Parts & \textcolor{teal}{Equipment}, \textcolor{teal}{IntakeEquip}, \textcolor{teal}{equip} & Adding parts via equipping & Certification  \\
     \hlhref{https://eips.ethereum.org/EIPS/eip-6381}{6381} &	Public Non-Fungible Token Emote Repository & \textcolor{teal}{emote}, \textcolor{teal}{emoteCountOf} & Enabling emote repository & Feedback system  \\
     \hlhref{https://eips.ethereum.org/EIPS/eip-6454}{6454} &	Minimal Transferable NFT detection interface & \textcolor{teal}{isTransferable} & Identifying transferability &    \\
     \hlhref{https://eips.ethereum.org/EIPS/eip-6672}{6672} &	Multi-redeemable NFTs & \textcolor{teal}{redeem}, \textcolor{teal}{getRedemptionIds}  & Enabling redemption & Financial market \\
     \hlhref{https://eips.ethereum.org/EIPS/eip-6808}{6808} &	Fungible Key Bound Token & \textcolor{teal}{addBindings}, allow (\textcolor{teal}{Transfer/Approval}) & Upgrading security to fungible & Financial market   \\
     \hlhref{https://eips.ethereum.org/EIPS/eip-6809}{6809} &	Non-Fungible Key Bound Token & \textcolor{teal}{addBindings}, allow (\textcolor{teal}{Transfer/Approval})& Upgrading security to non-fungible & Financial market \\

   \midrule 
   \hlhref{https://eips.ethereum.org/EIPS/eip-5008}{5008} &  ERC-721 Nonce Extension & \textcolor{teal}{nonce} & Adding nonce &   \\
   \hlhref{https://eips.ethereum.org/EIPS/eip-5114}{5114} &  Soulbound Badge &  \textcolor{teal}{CollectionUri}, \textcolor{teal}{badgeUri} & Bound to one-off transferring & Soulbound Items   \\
   \hlhref{https://eips.ethereum.org/EIPS/eip-5216}{5216} &  EIP-1155 Approval By Amount Extension & \textcolor{teal}{approve}, \textcolor{teal}{allowance} & Upgrading approval functions &  \\
   \hlhref{https://eips.ethereum.org/EIPS/eip-5496}{5496} &  Multi-privilege Management NFT Extension & \textcolor{teal}{setPrivilege}, \textcolor{teal}{privilegeExpires} & Adding shareable privileges & Marketplace  \\
   \hlhref{https://eips.ethereum.org/EIPS/eip-5585}{5585} &  ERC-721 NFT Authorization & \textcolor{teal}{authorizeuser}, \textcolor{teal}{transferUserRights} & Enabling cross-users authorization & Authorization  \\
   \hlhref{https://eips.ethereum.org/EIPS/eip-6982}{6982} &  Efficient Default Lockable Tokens & \textcolor{teal}{locked} & Enabling locks to reduce gas consumption &  \\
   \hlhref{https://eips.ethereum.org/EIPS/eip-7066}{7066} &  Lockable Extension for ERC-721 & \textcolor{teal}{lock}, \textcolor{teal}{transferAndLock} & Enabling locks to management trading & Authorization   \\
   \hlhref{https://eips.ethereum.org/EIPS/eip-7160}{7160} &  ERC-721 Multi-Metadata Extension & \textcolor{teal}{tokenURIs}, \textcolor{teal}{pinTokenURI} & Upgrading metadata URIs & File system   \\
   \hlhref{https://eips.ethereum.org/EIPS/eip-7231}{7231} &  Identity-aggregated NFT & \textcolor{teal}{setIdentitiesRoot}, \textcolor{teal}{verifyIdentitiesBinding} & Integrating Web2 and Web3 identities & Metaverse  \\
   \midrule 
    \hlhref{https://eips.ethereum.org/EIPS/eip-4973}{4973}  &  Account-bound Tokens  & \textcolor{teal}{unequip}, \textcolor{teal}{give}, \textcolor{teal}{take} & Bound to an account  & Game \\
    \hlhref{https://eips.ethereum.org/EIPS/eip-5521}{5521} &  Referable NFT  & \textcolor{teal}{referringOf}, \textcolor{teal}{setNodeReferring}  & Enabling reference relationship & Literature market \\
    \hlhref{https://eips.ethereum.org/EIPS/eip-6065}{6065}  & Real Estate Token   & \textcolor{teal}{debtOf}, \textcolor{teal}{managerOf}, \textcolor{teal}{geoJsonOf}, \textcolor{teal}{legalOwnerOf} & Adding real estate functions & Real estate market   \\
    \hlhref{https://eips.ethereum.org/EIPS/eip-6120}{6120}  & Universal Token Router  & \textcolor{teal}{exec}, \textcolor{teal}{pay}, \textcolor{teal}{discard} & Upgrading tranfer calls to reduce gas consumption &   \\
    \hlhref{https://eips.ethereum.org/EIPS/eip-6551}{6551} & 	Non-fungible Token Bound Accounts  & \textcolor{teal}{account}, \textcolor{teal}{token}, \textcolor{teal}{state} & Account bound to tokens &  \\
    \hlhref{https://eips.ethereum.org/EIPS/eip-6997}{6997} & 	ERC-721 with transaction validation step  & \textcolor{teal}{ValidateTransfer}, \textcolor{teal}{ValidateApproval} & Upgrading security to non-fungible & Financial market  \\
    \hlhref{https://eips.ethereum.org/EIPS/eip-7432}{7432} & 	Non-Fungible Token Roles  & \textcolor{teal}{grantRoleFrom}, \textcolor{teal}{roleExpirationDate} & Adding expirable role management & Rental market  \\
    
   \midrule 
   
   \hlhref{https://eips.ethereum.org/EIPS/eip-998}{998}   & Composable Non-Fungible Token &  \textcolor{teal}{rootOwnerOf}, \textcolor{teal}{transferChild} & Enabling composability between tokens & Metaverse, Game, Marketplace  \\
   \hlhref{https://eips.ethereum.org/EIPS/eip-4883}{4883}  & Composable SVG NFT & \textcolor{teal}{renderTokenById} & Extending composability to SVG tokens & Marketplace\\
   
   \hlhref{https://eips.ethereum.org/EIPS/eip-5173}{5173}  &	NFT Future Rewards & \textcolor{teal}{releaseFR}, \textcolor{teal}{retrieveFRInfo} & Enabling future earnings & Financial market \\
   \hlhref{https://eips.ethereum.org/EIPS/eip-5700}{5700}  &	Bindable Token Interface & \textcolor{teal}{bind}, \textcolor{teal}{boundBalanceOf} & Tokens bound to other tokens & Game, Rental market \\
   \hlhref{https://eips.ethereum.org/EIPS/eip-5727}{5727}  &	Semi-Fungible Soulbound Token & \textcolor{teal}{issue}, \textcolor{teal}{revoke}, \textcolor{teal}{verify} & Additional attribute for semi-fungible soulbound token & Soulbound Items \\
   \hlhref{https://eips.ethereum.org/EIPS/eip-5791}{5791}  &	Physical Backed Tokens & \textcolor{teal}{transferTokenWithChip} & Linking physical items & IoT industry \\
   
   \hlhref{https://eips.ethereum.org/EIPS/eip-6604}{6604}  &	Abstract Token & \textcolor{teal}{AbstractTokenMessage} & Abstracting tokens & Identification, Access credentials\\
   \hlhref{https://eips.ethereum.org/EIPS/eip-6682}{6682}  &	NFT Flashloans & \textcolor{teal}{flashFeeToken}, \textcolor{teal}{availableForFlashLoan} & Enabling flashloans & Real Estate, Financial market\\
   \hlhref{https://eips.ethereum.org/EIPS/eip-6785}{6785}  &	ERC-721 Utilities Information Extension & \textcolor{teal}{setUtilityUri}, \textcolor{teal}{utilityHistoryOf} & Adding utility to tokens & \\
   \hlhref{https://eips.ethereum.org/EIPS/eip-6786}{6786}  &	Registry for royalties payment for NFTs & \textcolor{teal}{payRoyalties}, \textcolor{teal}{getPaidRoyalties} & Upgrading royalties& Royalty payments \\
   \hlhref{https://eips.ethereum.org/EIPS/eip-6806}{6806}  &	ERC-721 Holding Time Tracking & \textcolor{teal}{getHoldingInfo}, \textcolor{teal}{\_afterTokenTransfer} & Adding holding time& Rental market \\
   \hlhref{https://eips.ethereum.org/EIPS/eip-6823}{6823}  &	Token Mapping Slot Retrieval Extension & \textcolor{teal}{getTokenLocationRoot} & Enhancing precision of off-chain transaction simulations &  \\
   \hlhref{https://eips.ethereum.org/EIPS/eip-6956}{6956}  &	Asset-bound Non-Fungible Tokens & \textcolor{teal}{tokenByAnchor}, \textcolor{teal}{OracleUpdate} & Bound to assets authorized by oracle & Digital twin \\

   \hlhref{https://eips.ethereum.org/EIPS/eip-7303}{7303}  &	Token-Controlled Token Circulation & \_grantRoleBy (\textcolor{teal}{ERC721, ERC1155}) & Enabling token-controlled token circulation & Authorization\\
   \hlhref{https://eips.ethereum.org/EIPS/eip-7507}{7507}  &	Multi-User NFT Extension & \textcolor{teal}{userExpires}, \textcolor{teal}{setUser} & Bound to multiple users & Rental market \\
   \hlhref{https://eips.ethereum.org/EIPS/eip-7015}{7015}  &	NFT Creator Attribution & \textcolor{teal}{\_validateSignature}, \textcolor{teal}{\_isValid} & Securing creator attribution & Authentication, Authorization \\
   \hlhref{https://eips.ethereum.org/EIPS/eip-7085}{7085}  &	NFT Relationship Enhancement & \textcolor{teal}{setRelationship}, \textcolor{teal}{setAttribute} & Enhancing reference relationship &  Literature market\\

   \hlhref{https://eips.ethereum.org/EIPS/eip-7496}{7496}  & NFT Dynamic Traits &  \textcolor{teal}{getTraitValue}, \textcolor{teal}{TraitUpdated} & Enabling trait values  &  Marketplace \\
   \hlhref{https://eips.ethereum.org/EIPS/eip-7498}{7498}  & NFT Redeemables & \textcolor{teal}{getCampaign}, \textcolor{teal}{redeem} & Extending redeemablility  & Marketplace  \\
   \hlhref{https://eips.ethereum.org/EIPS/eip-7510}{7510}  & Cross-Contract Hierarchical NFT & \textcolor{teal}{parentTokenOf}, \textcolor{teal}{setParentTokens} & Enabling cross-contracts hierarchy   & Metaverse  \\
   \hlhref{https://eips.ethereum.org/EIPS/eip-7513}{7513}  & Smart NFT - A Component for Intent-Centric & \textcolor{teal}{execute}, \textcolor{teal}{validatePermission} & Enabling executable NFTs  & Metaverse  \\
   \hlhref{https://eips.ethereum.org/EIPS/eip-7531}{7531}  &	Staked ERC-721 Ownership Recognition & \textcolor{teal}{rightsHolderOf} & Adding staking interface for Legacy NFTs  &  Game \\
   \hlhref{https://eips.ethereum.org/EIPS/eip-7548}{7548}  &	Open IP Protocol built on NFTs  & \textcolor{teal}{isParent}, \textcolor{teal}{isSibling}, \textcolor{teal}{Component} & Merging multiple IPs  & Rental market  \\
   \hlhref{https://eips.ethereum.org/EIPS/eip-7590}{7590}  &	ERC-20 Holder Extension for NFTs  & \textcolor{teal}{TransferERC20ToToken}, \textcolor{teal}{...From...} & Extending exchange of different token types  & Marketplace  \\

   \bottomrule

\end{tabular}
 }
\end{table*}

\end{document}